\documentclass[preprint2]{aastex}
\usepackage{rotating,times,pictex,graphicx,latexsym,color,pifont}
\usepackage{longtable}

\shorttitle{Which are the youngest protostars?}
\shortauthors{D.~Froebrich}

\begin{document}

\title{Which are the youngest protostars? \\ Determining properties of
confirmed and candidate Class\,0 sources by broad-band photometry}

\author{D.~Froebrich}
\affil{Dublin Institute for Advanced Studies, 5 Merrion Square, Dublin 2,
Ireland} 

\begin{abstract}

We searched the literature to obtain a complete list of known Class\,0 sources.
A list of 95 confirmed or candidate objects was compiled. To the best of our
knowledge, all published broad-band observations from 1\,$\mu$m to 3.5\,mm have
been collected and are assembled in a catalogue. These data were used to
determine physical properties (T$_{\rm bol}$, L$_{\rm bol}$, L$_{\rm
smm}$/L$_{\rm bol}$, M$_{\rm env}$) and for a uniform classification. 50
sources possess sufficient observational data and are classified as Class\,0 or
Class\,0/1 objects. The source properties are compared with different
evolutionary models to infer ages and masses, and their correlations are
investigated. About 25\% of the sources are found to be in a quiet accretion
phase or possess a significantly different time evolution of the accretion rate
than the average. In Taurus, with its isolated star formation mode, this seems
especially to be the case.

\end{abstract}



\section{Introduction}

Protostars begin their life soon after the collapse of their parental cloud
core begins. Such stellar embryos have masses of less than 10$^{-2}$\,M$_\odot$
(Larson \cite{l03}) and are deeply embedded in a dense and massive envelope of
gas and dust. The central object grows in mass by accreting material from this
envelope either by direct infall or via a flattened circumstellar disc. As long
as the central object is less massive than the surrounding envelope it is
called a Class\,0 object. Otherwise it is a Class\,1 protostar. Since this mass
ratio is not directly observable, other criteria have to be adopted for a
classification. Andr\'e et al. \cite{2000prpl.conf...59A} defined three
observational properties needed to classify an object as a Class\,0 source: (i)
an internal heating source (compact centimeter radio continuum) or bipolar
outflow as an indication of a central young stellar object; (ii) extended and
centrally peaked sub-millimeter continuum emission, indicating the spheroidal
envelope; (iii) a high ratio of sub-millimeter to bolometric luminosity
(L$_{\rm smm}$/L$_{\rm bol}$\,$>$\,0.005), with L$_{\rm smm}$ measured longward
of 350\,$\mu$m. It is shown in Andr\'{e} et al. \cite{1993ApJ...406..122A} that
this last criterion is equivalent to the mass ratio of the envelope and the
central star being larger than one. Other authors use the bolometric
temperature (T$_{\rm bol}$) instead of L$_{\rm smm}$/L$_{\rm bol}$ (e.g. Chen
et al. \cite{1995ApJ...445..377C, 1997ApJ...478..295C} use T$_{\rm
bol}$\,$<$\,70\,K).

The Class\,0 stage can be characterised as the main mass accretion phase, in
which the forming star gains the bulk of its final mass. This process is not
yet well understood. How does the mass accretion rate depend on the age and the
properties of the cloud, and is it governed by turbulence or ambipolar
diffusion? What is the lifetime of the Class\,0 sources? How are the source
properties connected to the molecular outflow, that is inevitably driven by
these objects? The answers to these questions will finally help us to
understand the feedback of the star formation process to the surrounding
molecular cloud and the initial mass function.

To seriously address these questions we need a large, homogeneously classified
sample of these very young protostellar objects. Several samples have been
compiled in the recent years (e.g. Chen et al. \cite{1995ApJ...445..377C},
Andr\'e et al. \cite{2000prpl.conf...59A}, Shirley et al.
\cite{2000ApJS..131..249S}, Motte \& Andr\'{e} \cite{2001A&A...365..440M}). All
these works have specific problems: (1) they concentrate only on a limited
number of sources; (2) different criterias for classification of Class\,0
objects are used; (3) only parts of the available observational data are
considered. 

The first step to obtain a large, homogeneously classified sample of Class\,0
sources is a complete search of the literature for such objects. Here we
present an as far as possible complete list of the youngest known protostars to
date. We will classify them by means of their spectral energy distribution
(SED) using all (to the best of our knowledge) published broad-band photometric
data. We derive basic properties of these sources (T$_{\rm bol}$, L$_{\rm
bol}$). The sub-mm data are used to estimate envelope masses (M$_{\rm env}$)
and sub-mm slopes of the SED ($\beta$). By comparing T$_{\rm bol}$, L$_{\rm
bol}$, and M$_{\rm env}$ with evolutionary models we determine ages and masses
of these sources and investigate limitations of the current observational data
and evolutionary models. This is of particular interest for forthcoming
powerful telescopes like Spitzer and ALMA, which will enable us to improve our
knowledge of these sources significantly. Given this expected growth in
observational data in the near future, the present catalogue might be quickly
outdated. Hence, all the data will be available on a
webpage\footnote{http://www.dias.ie/protostars} and updated regularly.

In Sect.\,\ref{sample} we introduce the source sample. The method of our data
analysis is then put forward in Sect.\,\ref{analysis}, including a description
of the source classification in Sect.\,\ref{classification}. Finally the
results and implications are discussed in Section\,\ref{discussion}.


\section{Source sample and broad band photometry}

\label{sample}

To obtain an as far as possible complete list of known Class\,0 objects we
combined the samples of candidates and sources mentioned in various
publications. In particular we combined: (1) sources with T$_{\rm
bol}$\,$<$\,100\,K from Chen et al. \cite{1995ApJ...445..377C}; (2) objects
with T$_{\rm dust}$\,$<$\,100\,K from Hurt et al. \cite{1996ApJ...460L..45H};
(3) Class\,0 sources listed in Bontemps et al. \cite{1996A&A...311..858B}; (4)
objects of type CL0 in Saraceno et al. \cite{1996A&A...309..827S}; (5)
candidates with T$_{\rm bol}$\,$<$\,100\,K from Chen et al.
\cite{1997ApJ...478..295C}; the source B\,35 could not be identified in SIMBAD;
(6) Objects with L$_{\rm bol}$/L$_{\rm smm}$\,$<$\,200 in Chini et al.
\cite{1997ApJ...474L.135C} with VLA detected counterparts in Reipurth et al.
\cite{1999AJ....118..983R}; (7) the Class\,0 sources in CB\,68 and CB\,232
mentioned in Huard et al. \cite{1999ApJ...526..833H}; (8) objects classified as
Class\,0 in Park et al. \cite{1999ApJ...520..223P}; (9) all objects listed in
Andr\'{e} et al. \cite{2000prpl.conf...59A}; (10) Class\,0 sources in Shirley
et al. \cite{2000ApJS..131..249S}; (11) sources indicated as Class\,0 or
Class\,0/1 in Motte \& Andr\'{e} \cite{2001A&A...365..440M}; (12) objects with
L$_{\rm FIR}$/L$_{\rm smm}$\,$<$\,200 in Chini et al.
\cite{2001A&A...369..155C}; (13) candidates with T$_{\rm bol}$\,$<$\,100\,K in
Lehtinen et al. \cite{2001A&A...367..311L}; (14) Class\,0 sources from Visser
et al. \cite{2002AJ....124.2756V}; (15) objects with T$_{\rm bol}$\,$<$\,100\,K
listed in Young et al. \cite{2003ApJS..145..111Y}; (16) Class\,0 objects and
candidates investigated in Froebrich et al. \cite{2003MNRAS.346..163F}; (17)
candidates and confirmed Class\,0 sources from Rengel et al.
\cite{2004A&A..inprep.R}; (18) the possible Class\,0 sources NGC\,7129\,FIRS2
(Eiroa et al. \cite{1998A&A...335..243E}), NGC\,2068\,LBS\,17 (Gibb \& Little
\cite{2000MNRAS.313..663G}), IC\,348\,MMS (Eisl\"offel et al.
\cite{2003ApJ...595..259E}), NGC\,7538\,S (Sandell et al.
\cite{2003ApJ...590L..45S}), and MonOB1\,IRAS12\,S1 (Wolf-Chase et al.
\cite{2003MNRAS.344..809W}). All investigated sources with their positions,
adapted distances, and references are listed in Appendix\,\ref{sourcesample} in
Table\,\ref{sources}. This table will be only available in electronic form and
lists the object name used in this paper and other common names of the sources.

Sources in the combined sample were identified by their position in the SIMBAD
database in order to avoid confusion due to unclear or conflicting
nomenclature. A search in the related publications given by SIMBAD for the
sources was performed and to our best knowledge all available broad-band
photometric data were extracted. We restricted the search for data ranging from
the near infrared (NIR) to wavelengths shorter than 3.5\,mm. At longer
wavelengths the emission of the envelope might be a combination of dust
continuum and free-free emission from the central source and may influence the
analysis of the envelope properties. NIR and mid-infrared detections with
brightness given in magnitudes were converted into Jansky using flux
zero-points of Wamsteker \cite{1981A&A....97..329W}. In case of non-existing
NIR detections in the literature we searched the 2MASS catalogue. If the source
was not detected by 2MASS, we determined an upper limit (the faintest flux in
the 2MASS catalogue within 200\arcsec\, of the source).


\section{Data analysis}

\label{analysis}

\subsection{Basic parameters}

The observed SEDs have to be characterised in order to be able to compare them
and learn about the object properties. We computed T$_{\rm bol}$ following the
method described in Chen et al. \cite{1995ApJ...445..377C}. This is the
temperature of a black body, which posseses the same mean frequency as the
respective source. The bolometric luminosity is determined by integrating the
SED and adapting a distance. Note that L$_{\rm bol}$ might be underestimated by
this method, when the emission maximum is not well covered (no data from 100 to
350\,$\mu$m). We compute the ratio of bolometric to sub-mm ($\lambda >
350$\,$\mu$m) luminosity. This ratio is used in the source classification (see
Sec.\,\ref{classification}). 

The mean spectral index $\alpha$ was determined by a powerlaw fit to fluxes
measured longward of 400\,$\mu$m. $\alpha$ is converted to the powerlaw index
of the dust emissivity (assuming optically thin emission) by
$\beta$\,=\,$\alpha$-2. The powerlaw fit allows for a uniform determination of
the flux density at 1.3\,mm, even if the object was not observed at this
wavelength. The fluxes are scaled to a distance of 300\,pc and transformed into
envelope masses by M$_{\rm env}$\,=\,1.5\,M$_\odot$\,*\,F$_{\rm 1.3mm}^{\rm
300pc}$[Jy] (following Motte and Andr\'e \cite{2001A&A...365..440M}). We used a
dust opacity at 1.3\,mm of 0.01\,cm$^2$\,g$^{-1}$ and a dust temperature of
24\,K for all objects. A lower dust temperature of 15\,K, as e.g. suggested by
Motte \& Andr\'e \cite{2001A&A...365..440M} for the Taurus sources, would lead
to envelope masses which are a factor of 1.6 higher.

The data for some sources are inhomogeneous. Different aperture sizes are used
at different wavelengths. To ensure a consistent treatment of the data for all
sources, datapoints were selected or excluded manually before the determination
of the source properties. In Table\,\ref{notes} 
(Appendix\,\ref{notesonsources}, available online only) we list datapoints not
used for the determinations, together with further notes. The obtained
parameters for all our sources are listed in Table\,\ref{param}. This table is
divided in three sections. First we list all sources where the SED is very well
sampled and all parameters can be determined accuratly. The second part
contains sources where the SED is only well determined on one side of the peak
and for the other side only upper limits (e.g. VLA\,1623) or very few data
(e.g. CB\,232) are available, hence mostly limits for the source properties
could be derived. In the last part we list sources that have very insufficient
data and where at best only $\beta$ and M$_{\rm env}$ or a lower limit for
L$_{\rm bol}$ can be deterimed. The objects Trifid-TC\,3, GF\,9-2, and
IC\,348\,MMS are not listed in this table since none of the source properties
could be determined.

\subsection{Source Classification}

\label{classification}

To classify the investigated objects we conducted three different tests for
objects listed in the first two parts of Table\,\ref{param}. A source was
negatively tested to be a Class\,0 object when: (1) T$_{\rm bol}$ is larger
than 80\,K; (2) L$_{\rm smm}$/L$_{\rm bol}$ is less than 0.005; (3) There is a
NIR ($\lambda$\,$<$5\,$\mu$m) detection. If all three tests are positive then
the source was marked as Class\,0 object (0). If at least two tests are
positive and the third could not be conducted or was negative, then the source
is marked as borderline object (0/1). In case of two negative tests the source
is classified as Class\,1 (1). For all remaining objects a classification
cannot be performed and the source is marked by '?'. In principle we further
need to test if the object possesses an extended envelope. Hence, we marked all
sources with a $^\dagger$ symbol, where non or no conclusive sub-mm or
millimeter map is available. The lack of such data mostly effects southern
hemisphere objects, which anyway could not be properly classified by our tests.

\begin{figure}[t]
\caption{\label{t_l} Distribution of L$_{\rm bol}$ (filled circles) and M$_{\rm
env}$ (open circles) against T$_{\rm bol}$ of the Class\,0
and Class\,0/1 sources with distances of less than 500\,pc.}
\includegraphics[width=7.cm, height=5.5cm, bb=90 350 310 520]{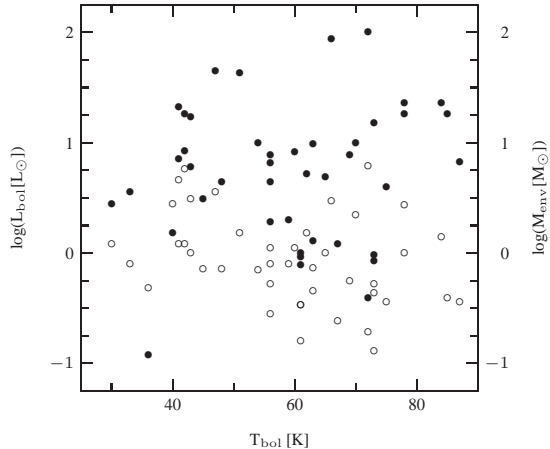}
\end{figure}

The classification obtained for each source according to these criteria is
listed in the last column of Table\,\ref{param}. Positions in the T$_{\rm
bol}$-L$_{\rm bol}$ diagram of all sources properly classified as Class\,0 or
Class\,0/1 can be seen in Fig.\,\ref{t_l} together with the T$_{\rm
bol}$-M$_{\rm env}$ positions. In order to obtain a well defined sample with as
few as possible selection effects, we excluded sources with distances larger
than 500\,pc in the diagram.

\subsection{Evolutionary Models}

\label{evolution}

Bolometric temperature, luminosity and envelope mass are three observational
quantities available for these young sources. To investigate physical
properties, such as final masses and ages, we need to compare these values with
evolutionary models. For the comparison we selected all sources from part one
in Table\,\ref{param} and the objects from part two with a clear classification
(0, 0/1, or 1), and where T$_{\rm bol}$, L$_{\rm bol}$, and M$_{\rm env}$ could
be determined. 

There are several evolutionary schemes available in the literature. We use
three different models to estimate ages and final masses of our sources. (1)
The evolutionary diagram presented in Fig.\,12 of Myers et al.
\cite{1998ApJ...492..703M}, where the infall rate matches the isothermal sphere
infall solution at early times and exponentially declines later. (2)
Evolutionary tracks shown in Fig.\,6b in Andr\'e et al.
\cite{2000prpl.conf...59A}, where accretion rate and envelope mass decline
exponentially. (3) The evolutionary scheme developed by Smith
\cite{1998Ap&SS.261..169S, 2000IrAJ...27...25S, s02}, where the mass accretion
rate first increases exponentially and then shows a powerlaw fall off. 

All of these evolutionary models are subject of shortcommings. They are based
on assumptions about the time dependence of the mass accretion rates. While the
overall decline with time has substantial observational support (Calvet et al.
\cite{2000prpl.conf..377C}), the particular behaviour is unclear. Analytical
and numerical models reproduce both, exponential decline (Shu et al.
\cite{2004ApJ...601..930S}) or well defined early peaks followed by a fall off
(Schmeja \& Klessen \cite{2004A&A...419..405S}). The usage of T$_{\rm bol}$ and
L$_{\rm bol}$ as input parameters (Smith \cite{1998Ap&SS.261..169S,
2000IrAJ...27...25S, s02}, Myers et al. \cite{1998ApJ...492..703M}) will
lead to uncertainties since both depend on the viewing angle, in contrast to
the envelope mass (Andr\'e et al. \cite{2000prpl.conf...59A}), which is
inferred from optically thin dust emission. Hence, the absolute values of the
determined ages and masses are model dependent but the relative values can be
used to put these youngest known protostars in an evolutionary sequence.
Calculated ages and final masses from all models are listed in
Table\,\ref{evolmod}.


\section{Discussion}

\label{discussion}

\subsection{Source classification}

We investigated SED's of 95 young protostellar sources. 59 of these objects
(60\,\%) possess sufficient observational data to accurately determine all
properties (T$_{\rm bol}$, L$_{\rm bol}$, M$_{\rm env}$). Out of these, 27
objects could be unambiguously classified as Class\,0 object, 23 as Class\,0/1,
and 9 as Class\,1 protostars (according to our definition). There is a
remaining relatively large number (36 objects; 40\,\%) where we do not have
sufficient information about the SED for a proper classification.

There are two main gaps in the observational data which mean this large
fraction could not be classified properly. (1) Only data at wavelengths shorter
than the emission maximum and non or only insufficient sub-mm or millimeter
observations are available. Hence, only upper limits for T$_{\rm bol}$ and
lower limits for L$_{\rm bol}$ can be determined. (2) Only data at wavelengths
longer than the maximum and no observations or just upper limits at shorter
wavelengths are available. For these objects only $\beta$ and M$_{\rm env}$ can
be estimated properly.

\subsection{Uncertainties}

All determined object parameters are subject to different errors, due to their
calculation. The powerlaw index of the opacity is determined from the mean
slope of the SED at $\lambda$\,$>$\,400\,$\mu$m. Using fluxes taken with
different aperture sizes to determine $\beta$ will lead to different results.
Estimating $\beta$ as the mean slope ensures that for each object a 'mean'
aperture size for the (sub)mm data is used. Typical errors of the estimated
values are 0.3. 

The luminosity is determined by integrating the SED and adopting a distance. An
error of 20\% for the distance (a typically value for most of the objects with
d\,$<$\,500\,pc) leads to 40\% uncertainty for the luminosity. Also the
luminosity might be underestimated when the SED is not well sampled at the
emission peak, or when huge parts of the SED are only described by upper limits
(e.g. VLA\,1623 or HH\,24\,MMS). Additional uncertainties are expected in
cluster regions, where current far-infrared instruments do not possess
sufficient angular resolution to separate individual sources (e.g. SVS\,13\,B).
Non-detections or observations in the NIR or at millimeter wavelength have
almost no influence, since most energy is radiated near the emission maximum.
All together the determined source luminosities are uncertain by about 50\% in
most cases. 

For the bolometric temperature, datapoints at short wavelengths are important.
In the case of well sampled SEDs from the NIR to the sub-mm, T$_{\rm bol}$ can
be determined with about 5 to 10\,K accuracy. In the case of upper limits in
the NIR, the same accuracy applies as long as the limits are at least 5-6
orders of magnitude lower than the flux at the maximum of the emission. Note
that the sources classified as Class\,0/1 have typically NIR detections 4-5
orders of magnitude below the emission maximum, while for the Class\,0 sources
we find typical non-detections 5-7 orders of magnitude below the maximum. In
case of 'bad' NIR limits the uncertainty in T$_{\rm bol}$ can well exceed
10\,K. Hence for our Class\,0 sources the temperatures are accurate to 10\,K,
while for the Class\,0/1 objects they are certain to 20\,K. Note that for older
sources (Class\,0/1 or Class\,1) viewing angle effects might become important.
They lead in case of edge-on sources to underestimates of T$_{\rm bol}$ and
L$_{\rm bol}$. Subsequently the L$_{\rm smm}$/L$_{\rm bol}$ ratio is
overestimated, influencing the source classification.

The envelope mass is also subject of various sources of errors. Beside the
problems with measuring the fluxes in different apertures, the distance is used
in the determination. Further the applied dust opacities are uncertain by a
factor of two (e.g. Motte \& Andr\'e \cite{2001A&A...365..440M}), and the dust
temperatures are not very well known. Hence, the masses might suffer from
errors of a factor of three. 

How do these errors influence inferred ages and masses? The models of Smith
\cite{1998Ap&SS.261..169S, 2000IrAJ...27...25S, s02} and Myers et al.
\cite{1998ApJ...492..703M} use T$_{\rm bol}$ and L$_{\rm bol}$ as main input
parameter. In both cases the error in the measured luminosity leads to
uncertainties of about a factor of two for the estimated final masses. Errors
in T$_{\rm bol}$ will transform into age uncertainties which are about 30\%.
The model of Andr\'e et al. \cite{2000prpl.conf...59A} uses the envelope masses
as input. Hence inferred final masses are uncertain by a factor of three and
estimated source ages are uncertain by a factor of two. Note that these
uncertainties do not take into account systematic errors due to the various
assumptions made in the different evolutionary models.

\subsection{Distribution of source parameters}

To investigate the statistical properties of the obtained Class\,0 sample, we
restrict all the following discussions to sources classified as Class\,0 or
Class\,0/1 objects within 500\,pc. 

The power-law index of the opacity is evenly distributed from 0.0 to 2.0 with a
broad maximum between 0.5 and 1.5. For our sources T$_{\rm bol}$ ranges from 30
to 90\,K, where the 90\,K limit is due to our classification criterion for
Class\,0 sources. The number of sources gradually rises from 30 to 70\,K and
falls off at higher temperatures. One expects a rising (or at least constant)
number of sources towards higher temperatures, because with higher temperature
an increased age and a slower evolution in temperature is predicted by the
models. Hence, the T$_{\rm bol}$ distribution shows that there are missing
objects in our sample in this temperature range. This might be due to a
classification of these objects as Class\,1 protostars and reflect the fact
that sources with the same bolometric temperatures are not necessarily in the
same evolutionary state. The distribution of L$_{\rm bol}$ corresponds very
well with the mass function (see below), due to the assumptions in the
evolutionary models. Similarly the distribution of M$_{\rm env}$ shows a peak
at about one solar mass. At lower masses our sample certainly suffers from
incompleteness. The L$_{\rm smm}$/L$_{\rm bol}$ ratio shows a broad
distribution between 0.01 and 0.1. Only very few larger values are found
(IRAM\,04191, L\,1448\,NW). 

We further investigated if the source properties are correlated. There is no
correlation between the dust opacity $\beta$ and the parameters T$_{\rm bol}$
and L$_{\rm bol}$ (correlation coefficients (c.c.) -0.034 and -0.102). A weak
correlation (c.c. -0.412) is found between $\beta$ and M$_{\rm env}$, but this
is due to a few high envelope mass sources. Also T$_{\rm bol}$ is not
correlated with L$_{\rm bol}$ and M$_{\rm env}$ (c.c. 0.149 and -0.219). On the
other hand the envelope mass shows a correlation with the bolometric luminosity
(c.c. 0.660). This correlation indicates that objects with more massive
envelopes possess a higher luminosity and hence higher accretion rates.

\subsection{The different evolutionary models}

Due to the different model assumptions about the time evolution of the mass
accretion rates, all three investigated evolutionary models lead to different
values for ages and final star masses. The ages especially are very model
dependent. We investigate if there are correlations between the inferred values
for age and final mass for the different evolutionary models in our Class\,0
and Class\,0/1 sample. 

On first sight the ages differ significantly between the models. Ages from
Myers et al. \cite{1998ApJ...492..703M} are much larger than usually assumed
for Class\,0 sources (a couple of 10$^4$\,yrs). In the model of Andr\'e et al.
\cite{2000prpl.conf...59A} the evolution in the first 10000\,yrs is very fast,
and hence the inferred ages are on average very small ($<$\,10$^4$\,yrs).
However, there is a weak correlation between the ages inferred from the model
of Smith \cite{1998Ap&SS.261..169S, 2000IrAJ...27...25S, s02} and Myers et al.
\cite{1998ApJ...492..703M} (c.c. 0.425). Comparing the models from Andr\'e et
al. \cite{2000prpl.conf...59A} with Myers et al. \cite{1998ApJ...492..703M} and
Smith \cite{1998Ap&SS.261..169S, 2000IrAJ...27...25S, s02} we find no
correlation (c.c. 0.143, 0.263). 

The predicted final masses are much more important than the ages, since the
mass distribution should represent the observed initial mass function. Here we
investigate if the relative masses obtained from the different models are
comparable. There are no obvious differences in the inferred masses between the
models. The models of Myers et al. \cite{1998ApJ...492..703M} and Andr\'e et
al. \cite{2000prpl.conf...59A} are limited in their mass range (0.3-0.7 and
0.2-3.0\,M$_\odot$, respectively). We find good correlations between masses
obtained from the different models: c.c.=0.676 for Smith
\cite{1998Ap&SS.261..169S, 2000IrAJ...27...25S, s02} and Myers et al.
\cite{1998ApJ...492..703M}; c.c.=0.656 for Smith \cite{1998Ap&SS.261..169S,
2000IrAJ...27...25S, s02} and Andr\'e et al. \cite{2000prpl.conf...59A} and
c.c.=0.771 for Myers et al. \cite{1998ApJ...492..703M} and Andr\'e et al.
\cite{2000prpl.conf...59A}). 

\subsection{The Mass Function}

The evolutionary scheme by Smith \cite{1998Ap&SS.261..169S,
2000IrAJ...27...25S, s02} allows the comparison of all three main observational
quantities (T$_{\rm bol}$, L$_{\rm bol}$, M$_{\rm env}$) with the model. Also
there is no restriction on the final mass as in the other models. Further, the
assumed mass accretion rates are very similar to accretion rates obtained by
hydrodynamical simulations of star formation (e.g. Klessen
\cite{2001ApJ...550L..77K}, Schmeja \& Klessen \cite{2004A&A...419..405S}).
Hence it is worth to investigate the resulting final mass function obtained for
our sources using this model. Certainly the absolute values for the masses are
uncertain, but giving the good correlations to the other models (see above),
the relative masses, needed to determine the slope in the mass function, are
correct.

\begin{figure}[t]
\caption{\label{imf} Mass Function for the properly classified Class\,0 and
Class\,0/1 objects within 500\,pc, obtained from the model of Smith
\cite{1998Ap&SS.261..169S, 2000IrAJ...27...25S, s02}.}
\includegraphics[width=7.cm, height=5.9cm, bb=90 340 310 530]{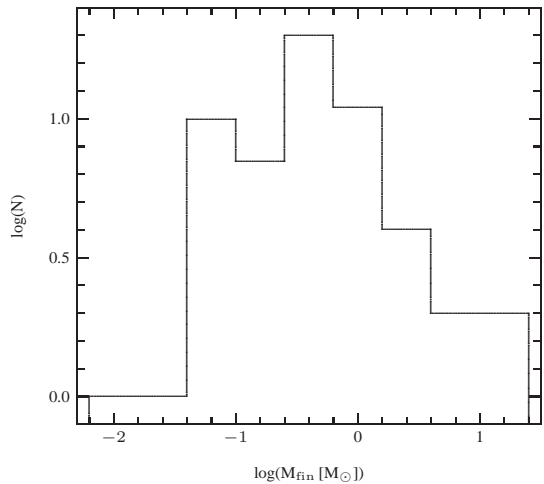}
\end{figure}

In Fig.\,\ref{imf} we present the resulting mass function for our sample of
Class\,0 and Class\,0/1 objects within 500\,pc. The slope determined for
objects with M\,$>$\,0.5\,M$_\odot$ is -0.9\,$\pm$\,0.2. Considering the low
number of sources, and that no binary correction is done, this is in good
agreement with the Salpeter slope of -1.35 measured for the solar neighborhood.
We estimate a completeness limit in our sample for objects with final masses
below 0.4\,M$_\odot$. 

There is, however, a number of predicted very low mass sources. These are
objects where the model of Smith \cite{1998Ap&SS.261..169S,
2000IrAJ...27...25S, s02} seems not to work properly. Looking in detail, we
find that the evolutionary scheme of Smith \cite{1998Ap&SS.261..169S,
2000IrAJ...27...25S, s02} can explain the all observed properties (T$_{\rm
bol}$, L$_{\rm bol}$, M$_{\rm env}$) for most of the sources simultaneously. In
case of eleven objects, however, the scheme is not working (IRAS\,03256+3055,
IRAS\,03282+3035, B\,213, IRAM\,04191, L\,1448\,N, IRAS\,04325+2402,
IRAS\,04368+2557, IRAS 15398-3359, HH\,24\,MMS, NGC\,2068\,LBS\,17, VLA\,1623).
These sources possess a much lower luminosity (considering their T$_{\rm bol}$
and M$_{\rm env}$ values) than the average of the objects. There are also three
objects where the luminosity is much above the average (NGC\,1333-I2,
HH\,212-MM, L\,1641\,N). Note that this applies also for the more distant
objects L\,1246-SMM\,1 (luminosity lower) and Cep\,E, IRAS\,20050+2720, and
MonOB1\,IRAS12\,S1 (luminosity higher). For the remaining sources the model
explains very well all three observed properties. 

There are two reasons for this inability to explain the data of all our
objects. (1) One might be found in the mass accretion rates obtained by Klessen
\cite{2001ApJ...550L..77K} and Schmeja \& Klessen \cite{2004A&A...419..405S}.
It turnes out that their accretion rates on average show a time evolution
similar to the one used in the model of Smith \cite{1998Ap&SS.261..169S,
2000IrAJ...27...25S, s02}. But on short timescales the accretion rate varies
significantly from this average. There are periods with lower and higher
accretion rates resulting in higher and lower luminosities. Hence, the very low
luminosity sources might represent a quiet accretion phase, while the higher
luminosity sources represent phases of enhanced accretion. (2) A second reason
might be that these sources do not accrete their mass in the same way. Hence,
the mass accretion rate shows a signifficantly different evolution in these
objects.

In case of the latter, this suggests two different groups of sources. 'Normal'
sources, possessing accretion rates similar to that obtained by Schmeja \&
Klessen \cite{2004A&A...419..405S}, and 'abnormal' sources showing on average
lower luminosities/accretion rates, considering their T$_{\rm bol}$ and M$_{\rm
env}$ values. Note that viewing angle effects could be responsible for some of
these objects. These low luminosity sources might represent objects where
ambipolar diffusion dominates the accretion process instead of turbulence. A
detailed analysis and discussion of this subject can be found in Froebrich et
al. \cite{2004A&A.inprep.F}.

\subsection{The different Star Forming Regions}

Does the star formation process, in particular the time dependence of the mass
accretion rate vary with the star forming region or is it a uniform function?
Due to the low number of sources a proper statistical analysis of this question
is not possible yet. However, in order to see possible trends we chose four
regions (Perseus, Taurus, Orion, and Serpens) where we have a sufficient number
of confirmed Class\,0 or Class\,0/1 objects (13, 4, 10, 6, respectively) and
compared these regions with the average of all sources.

In our sample of Class\,0 and Class\,0/1 sources about 25\% do not follow the
T$_{\rm bol}$-L$_{\rm bol}$-M$_{\rm env}$ relation of the majority of our
objects. These sources possess much lower luminosities than suggested by their
bolometric temperature and envelope mass. Hence, they might be in a phase of
lower mass accretion or are gaining their mass in a different way. Perseus
reflects the overall average with three out of 13 objects. The same applies for
Orion (two out of ten), where we might miss low luminosity objects due to the
larger distance. In Serpens all of the six sources show a 'normal' behaviour.
This is, according to the small number of objects, still in agreement with the
average. Taurus, however, shows a very different picture. Here all four
Class\,0 or Class\,0/1 objects fall into the low luminosity category. Objects
showing a 'normal' behaviour should have been easily detected since their
higher luminosity. Hence, even if there are just four sources this strongly
suggests that in a region of isolated star formation (as in Taurus) the time
evolution of the mass accretion rates is different from regions where stars
form in clusters (in agreement with previous findings from Henriksen et al.
\cite{1997A&A...323..549H}). Due to the bad statistics nothing can be said
about differences among the clusters.


\section{Conclusions}

A literature search for photometric broadband observations of 95 confirmed or
candidate Class\,0 sources was conducted. To our best knowledge all available
broadband photometric data was used to construct SEDs from 1\,$\mu$m to
3.5\,mm. If possible we determined basic properties of the sources (sub-mm
slope of the SED, T$_{\rm bol}$, L$_{\rm bol}$, L$_{\rm smm}$/L$_{\rm bol}$,
and M$_{\rm env}$). For 59 objects sufficient enough data are available for a
proper determination of the source parameters. 27 of these are classified as
Class\,0, 23 as Class\,0/1, and 9 as Class\,1 protostars. 

To investigate the statistical properties of the obtained sample of very young
protostars we used all objetcs within 500\,pc which are properly classified as
Class\,0 or Class\,0/1 to determine age and final star mass. Therefor we used
three different evolutionary models for protostars from Smith
\cite{1998Ap&SS.261..169S, 2000IrAJ...27...25S, s02}, Myers et al.
\cite{1998ApJ...492..703M}, and Andr\'e et al. \cite{2000prpl.conf...59A}.
Considering the uncertainties in the measured source properties the predicted
final star masses from the different models show a good correlation. The
absolute ages are, however, very model dependent.

An investigation of the final star masses and the resulting mass function shows
a good agreement with the IMF for stars with M\,$>$\,0.5\,M$_\odot$. Our
obtained Class\,0 sample is limited to sources with final masses above
0.4\,M$_\odot$. A number of objects (25\,\%) is found that possess a much lower
luminosity than the rest of our sample, considering their bolometric
temperature and envelope mass. These objects might be in a 'quiet' accretion
phase or the time evolution of their mass accretion rate is significantly
different from the majority. This might especially be the case in the Taurus
star forming region, where all identified Class\,0 or Class\,0/1 sources belong
to this group. 


\begin{acknowledgements}

We are grateful to M.\,Rengel for providing a preprint of her paper. We thank
the anonymous referee for very helpful comments and suggestions. M.D.\,Smith is
acknowledged for providing the code of his evolutionary model and his comments.
We thank  M.P.\,Redman, T.P.\,Ray, and \`A.\,Gras-Vel\'azquez for their help.
This publication makes use of the Protostars Webpage hosted by the Dublin
Institute for Advanced Studies. D.\,Froebrich received financial support by the
Cosmo-Grid project, funded by the Program for Research in Third Level
Institutions under the National Development Plan and with assistance from the
European Regional Development Fund. This research has made use of the SIMBAD
database, operated at CDS, Strasbourg, France. This research has made use of
NASA's Astrophysics Data System. This publication makes use of data products
from the Two Micron All Sky Survey, which is a joint project of the University
of Massachusetts and the Infrared Processing and Analysis Center/California
Institute of Technology, funded by the National Aeronautics and Space
Administration and the National Science Foundation.

\end{acknowledgements}



\onecolumn
\renewcommand{\arraystretch}{0.85}
\begin{center}
\begin{longtable}{l|ccccc|r}
\caption{\label{param} \bfseries Obtained object parameters, sorted by SED
quality}
\\ 
\noalign{\smallskip}
\hline
\noalign{\smallskip}
Object$^a$ & $\beta$ & T$_{\rm bol}$ & L$_{\rm bol}$ & L$_{\rm smm}$/L$_{\rm bol}$ 
                     & M$_{\rm env}$ & Class$^b$ \\
\noalign{\smallskip}
           &         & [K]       &[L$_\odot$]&  &   [M$_\odot$]        &       \\ 
\noalign{\smallskip}
\hline
\noalign{\smallskip}
\endfirsthead
\noalign{\smallskip}
\multicolumn{7}{c}{\bfseries \tablename\ \thetable{} -- continued from previous page} \\
\noalign{\smallskip}
\hline
\noalign{\smallskip}
Object$^a$ & $\beta$ & T$_{\rm bol}$ & L$_{\rm bol}$ & L$_{\rm smm}$/L$_{\rm bol}$ 
                     & M$_{\rm env}$ & Class$^b$ \\
\noalign{\smallskip}
           &         & [K]       &[L$_\odot$]&     &   [M$_\odot$]                &       \\ 
\noalign{\smallskip}
\hline
\noalign{\smallskip}
\endhead
\noalign{\smallskip}
\hline
\noalign{\smallskip}
\multicolumn{7}{r}{Continued on next page} \\
\noalign{\smallskip}
\endfoot
\noalign{\smallskip}
\hline
\noalign{\smallskip}
\endlastfoot
\noalign{\smallskip}
 L\,1448-I2                & 1.0   & 43      & 6.0     & 0.030    & 1.0   &  0/1 (yyn6)\\
 L\,1448\,NW               & 2.1   & $<$30   & $<$2.8  & $>$0.16  & 1.2   &  0   (yyy6)\\
 L\,1448\,N                & 0.6   & 70      & 10      & 0.028    & 2.2   &  0/1 (yyn4)\\
 L\,1448\,C                & 1.4   & $<$60   & 8.3     & 0.029    & 1.1   &  0   (yyy6)\\
 RNO\,15\,FIR              & 1.5   & 63      & 9.7     & 0.016    & 0.45  &  0/1 (yyn5)\\
 RNO\,15                   & 1.5   & $<$73   & 15      & 0.013    & 0.43  &  0   (yyy6)\\
 NGC\,1333\,I1             & 1.9   & $<$85   & 18      & 0.010    & 0.39  &  0/1 (?yy6)\\
 IRAS\,03256+3055          & 0.3   & $<$61   & 1.0     & 0.043    & 0.34  &  0$^\dagger$ (yyy5)\\
 NGC\,1333-I2              & 1.4   & $<$51   & 43      & 0.010    & 1.5   &  0   (yyy7)\\
 NGC\,1333-I4\,A           & 0.3   & $<$42   & 18      & 0.037    & 5.8   &  0   (yyy6)\\
 NGC\,1333-I4\,B           & 1.2   & $<$43   & 17      & 0.036    & 3.1   &  0   (yyy6)\\
 IRAS\,03282+3035          & 0.5   & $<$63   & 1.3     & 0.062    & 0.73  &  0   (yyy5)\\
 HH\,211-MM                & 1.3   & $<$33   & 3.6     & 0.046    & 0.80  &  0   (yyy6)\\
 B\,213                    & 0.7   & 72      & $<$0.39 & $<$0.028 & 0.19  &  0/1 (yyn5)\\
 IRAM\,04191               & 0.6   & $<$36   & 0.12    & 0.208    & 0.48  &  0   (yyy5)\\
 L\,1551-IRS\,5            & 1.2   & 92      & 22      & 0.008    & 1.6   &  1   (nyn4)\\
 L\,1551-NE                & 0.5   & 91      & 4.2     & 0.009    & 0.57  &  1   (nyn5)\\
 IRAS\,04325+2402          & 1.4   & 73      & 0.97    & 0.057    & 0.52  &  0/1 (yyn4)\\
 L\,1527                   & 1.6   & 56      & 1.9     & 0.056    & 0.80  &  0/1 (yyn4)\\
 IRAS\,05173-0555          & 1.0   & $<$87   & 6.7     & 0.014    & 0.36  &  0/1 (?yy5)\\
 RNO\,43-MM                & 0.7   & 56      & 6.5     & 0.018    & 0.52  &  0/1 (yyn5)\\
 OMC\,3-MM\,9              & 0.8   & 115     & 130     & 0.004    & 1.9   &  1   (nnn6)\\
 OMC\,3-MM\,7              & 1.2   & 123     & 53      & 0.008    & 1.2   &  1   (nyn4)\\
 L\,1641\,N                & 1.1   & 66      & 88      & 0.008    & 3.0   &  0/1 (yyn5)\\
 HH\,147\,MMS              & 0.9   & 208     & 45      & 0.002    & 0.39  &  1   (nnn3)\\
 HH\,212-MM                & 1.0   & $<$56   & 7.7     & 0.009    & 0.28  &  0   (yyy5)\\
 HH\,25\,MMS               & 1.7   & $<$41   & 7.2     & 0.063    & 1.2   &  0   (yyy5)\\
 HH\,111\,MMS              & 0.9   & 78      & 23      & 0.010    & 1.0   &  0/1 (yyn5)\\
 MonOB1\,IRAS12\,S1        & 1.3   & 41      & 76      & 0.045    & 5.5   &  0/1 (yyn5)\\
 NGC\,2264\,G-VLA\,2       & 1.1   & $<$70   & 13      & 0.026    & 0.61  &  0   (yyy5)\\
 IRAS\,15398-3359          & 1.7   & 61      & 0.92    & 0.048    & 0.34  &  0/1 (yyn4)\\
 IRAS\,16293-2422          & 0.8   & $<$41   & 21      & 0.019    & 4.6   &  0   (yyy7)\\
 IRAS\,16544-1604          & 1.0   & $<$67   & 1.2     & 0.019    & 0.24  &  0   (yyy5)\\
 IRAS\,18148-0440          & 1.1   & $<$54   & 10      & 0.010    & 0.70  &  0/1 (yyn5)\\
 Serp-S68\,N               & 1.4   & $<$56   & 4.4     & 0.060    & 1.1   &  0   (yyy5)\\
 Serp-FIRS\,1              & 0.6   & 47      & 45      & 0.011    & 3.6   &  0/1 (yyn6)\\
 Serp\,SMM\,5              & -0.6  & 113     & 3.6     & 0.018    & 1.4   &  1   (nyn4)\\
 Serp-SMM\,4               & 0.6   & $<$62   & 5.2     & 0.028    & 1.5   &  0   (yyy6)\\
 Serp-SMM\,3               & 0.5   & $<$65   & 4.9     & 0.019    & 1.0   &  0   (yyy5)\\
 Serp-SMM\,2               & 1.1   & $<$75   & 4.0     & 0.012    & 0.36  &  0   (yyy5)\\
 IRAS\,18331-0035          & 1.8   & 48      & 4.4     & 0.057    & 0.72  &  0/1 (yyn4)\\
 IRAS\,18148-0440          & 1.1   & 95      & 3.3     & 0.040    & 0.67  &  1   (nyn3)\\
 IRAS\,19345+0727          & 1.2   & $<$45   & 3.1     & 0.040    & 0.72  &  0   (yyy6)\\
 IRAS\,20050+2720          & 1.4   & 75      & 280     & 0.008    & 4.0   &  0/1 (yyn5)\\
 IRAS\,20386+6751          & 1.0   & $<$42   & 8.4     & 0.033    & 1.2   &  0   (yyy5)\\
 IRAS\,21017+6742          & 1.5   & 73      & 0.86    & 0.050    & 0.18  &  0/1 (yyn5)\\
 CB\,230                   & 1.0   & 69      & 7.7     & 0.022    & 0.56  &  0/1 (yyn4)\\
 L\,944-SMM\,1             & 0.8   & $<$80   & 4.7     & 0.035    & 0.44  &  0   (yyy5)\\
 L\,1251\,B                & 0.2   & 83      & 9.0     & 0.023    & 2.1   &  1   (nyn4)\\
 Cep\,E-MM                 & 2.0   & 56      & 89      & 0.027    & 2.4   &  0   (yyy8)\\
 L\,1246-SMM\,1            & 0.6   & $<$95   & 6.6     & 0.044    & 0.96  &  0/1 (?yy5)\\
 IRAS\,23238+7401          & 1.6   & $<$61   & 0.78    & 0.038    & 0.16  &  0   (yyy5)\\
 \noalign{\smallskip}
 \hline
 \noalign{\smallskip}
 SVS\,13\,B\,MMS\,1$^{*}$  & 1.1   & $<$163  & $<$82   & $>$0.004 & 1.2   &  ?   (??y4)\\
 SVS\,13\,B\,MMS\,2$^{*}$  & 1.3   & $<$165  & $<$81   & $>$0.003 & 1.1   &  ?   (??y4)\\
 SVS\,13\,B\,MMS\,3$^{*}$  & 1.4   & $<$169  & $<$79   & $>$0.002 & 0.42  &  ?   (??y4)\\
 HH\,114\,MMS$^{*}$        & 0.5   & $<$84   & $<$23   & $>$0.013 & 1.4   &  0/1 (?yy6)\\
 OMC\,3-MM\,6$^{*}$        & 0.7   & $<$72   & $<$100  & $>$0.010 & 6.1   &  0   (yyy7)\\
 NGC\,2023-MM\,1$^{*}$     & 0.2   & $<$87   & 200     & 0.002    & 2.8   &  ?   (?ny7)\\
 HH\,24\,MMS$^{*}$         & 0.1   & $<$78   & $<$18   & $>$0.019 & 2.7   &  0   (yyy5)\\
 NGC\,2068\,LBS\,17$^{*}$  & -0.7  & $<$40   & 1.5     & 0.107    & 2.8    &  0   (yyy5)\\
 IRAS\,08076-3556$^{*}$    & 0.0   & 117     & $>$10   & $<$0.009 & 1.1   &  1   (n?n5)\\
 Ced\,110\,IRS\,10$^{*}$   & ---   & 36      & 0.38    & 0.063    & ---   &  0$^\dagger$   (yyy5)\\
 Ced\,110\,IRS\,4$^{*}$    & ---   & 59      & 1.0     & 0.012    & ---   &  0/1$^\dagger$ (yyn5)\\
 BHR\,71-MM$^{*}$          & ---   & 68      & 7.9     & 0.008    & ---   &  0/1$^\dagger$ (yyn5)\\
 IRAS\,12553-7651$^{*}$    & ---   & 241     & 1.5     & 0.002    & ---   &  1$^\dagger$   (nnn3)\\
 IRAS\,13036-7644$^{*}$    & ---   & 77      & 0.94    & 0.011    & ---   &  0$^\dagger$   (yyy5)\\
 VLA\,1623$^{*}$           & 1.4   & $<$59   & $<$2.0  & $>$0.052 & 0.80  &  0   (yyy7)\\
 CB\,232$^{*}$             & ---   & 150     & 4.2     & 0.017    & ---   &  1   (nyn4)\\
 NGC\,7129\,FIRS\,2$^{*}$  & 0.9   & 72      & 330     & 0.011    & 8.6   &  0/1 (yyn4)\\
 L\,1211$^{*}$             & 1.7   & 191     & 62      & 0.005    & 0.28  &  1$^\dagger$   (nyn4)\\
 IRAS\,23385+6053$^{*}$    & 1.3   & $<$54   & 19000   & 0.002    & 12    &  0/1 (yny7)\\
 \noalign{\smallskip}
 \hline
 \noalign{\smallskip}
 W3\,OH$^{**}$             & 4.5   & ---     & ---     & ---      & 190   &  ---       \\ 
 L\,1634\,IRS\,7$^{**}$    & 1.5   & ---     & ---     & ---      & 0.60  &  ---       \\ 
 HH\,1-2\,MMS\,3$^{**}$    & 1.7   & ---     & ---     & ---      & 0.31  &  ---       \\ 
 HH\,1-2\,MMS\,2$^{**}$    & 2.3   & ---     & ---     & ---      & 0.19  &  ---       \\ 
 L\,1641-VLA\,1$^{**}$     & 0.8   & ---     & ---     & ---      & 1.6   &  ---       \\ 
 L\,1641\,SMS\,III$^{**}$  & 0.1   & ---     & ---     & ---      & 7.7   &  ---       \\ 
 L\,1641\,SMS\,IV$^{**}$   & 1.4   & ---     & ---     & ---      & 4.0   &  ---       \\ 
 NGC\,2024-FIR\,5$^{**}$   & 1.0   & ---     & ---     & ---      & 8.2   &  ---       \\ 
 NGC\,2024-FIR\,6$^{**}$   & 1.0   & ---     & ---     & ---      & 4.9   &  ---       \\ 
 IRAS\,12500-7658$^{**}$   & ---   & ---     & $>$0.62 & ---      & ---   &  ---       \\ 
 IRAS\,12533-7632$^{**}$   & ---   & ---     & $>$0.26 & ---      & ---   &  ---       \\ 
 IRAS\,12554-7635$^{**}$   & ---   & ---     & $>$0.17 & ---      & ---   &  ---       \\ 
 IRAS\,16017-3936$^{**}$   & ---   & ---     & $>$0.62 & ---      & ---   &  ---       \\ 
 IRAS\,16493-4242$^{**}$   & ---   & ---     & $>$4.8  & ---      & ---   &  ---       \\ 
 HH\,108\,MMS$^{**}$       & 1.2   & ---     & ---     & ---      & 0.49  &  ---       \\ 
 G\,34.24+0.13\,MM$^{**}$  & 0.9   & ---     & ---     & ---      & 2.8   &  ---       \\ 
 R\,CrA\,R1$^{**}$         & ---   & ---     & $>$3.5  & ---      & ---   &  ---       \\ 
 IRAS\,19180+1114$^{**}$   & ---   & ---     & $>$5.2  & ---      & ---   &  ---       \\ 
 S\,106-SMM$^{**}$         & 1.1   & ---     & ---     & ---      & 4.9   &  ---       \\ 
 IC\,1396\,W$^{**}$        & ---   & ---     & $>$26   & ---      & ---   &  ---       \\ 
 NGC\,7538\,S$^{**}$       & 2.2   & ---     & ---     & ---      & 51    &  ---       \\ 
\noalign{\smallskip}
\end{longtable}
\renewcommand{\arraystretch}{1.0}
\end{center}
\vspace{-0.75cm}
$^a$ Object name used in this paper. Further common names are given in
Table\,\ref{sources}. The $^{*}$ markes objects where the SED is is only well
determined on one side of the peak and for the other side only upper limits or
very few data is available. The $^{**}$ markes objects with too few
observational data to properly estimate even a limit for T$_{\rm bol}$. \\ 
$^b$ Classification due to the procedure described in
Sect.\,\ref{classification}. In brackets we give as 'y', 'n', or '?' the
answers for our three different tests if the object is a Class\,0 source. The
ending number gives the orders of magnitude between the maximum of the emission
and the NIR detection/upper limit. \\ 
$^\dagger$ Classification uncertain, since non or no conclusive sub-mm or 
millimeter maps are available. \\
\twocolumn

\onecolumn
\renewcommand{\arraystretch}{1.0}
\begin{center}
\begin{longtable}{l|r|ccc|ccc}
\caption{\label{evolmod} \bfseries Age and Mass estimates from evolutionary models} \\  
\noalign{\smallskip}
\hline
\noalign{\smallskip}
Object$^a$ & Class$^b$ & \multicolumn{3}{|c|}{Age$^c$\,[10$^3$\,yrs]} & 
                         \multicolumn{3}{|c}{M$^c_{\rm fin}$\,[M$_\odot$]} \\ 
\noalign{\smallskip}
 & & Smith & Myers & Andr\'e & Smith & Myers & Andr\'e \\
\noalign{\smallskip}
\hline
\noalign{\smallskip}
\endfirsthead
\noalign{\smallskip}
\multicolumn{6}{c}{\bfseries \tablename\ \thetable{} -- continued from previous
page} \\ 
\noalign{\smallskip}
\hline
\noalign{\smallskip}
Object$^a$ & Class$^b$ & \multicolumn{3}{|c|}{Age$^c$\,[10$^3$\,yrs]} & 
                         \multicolumn{3}{|c}{M$^c_{\rm fin}$\,[M$_\odot$]} \\ 
\noalign{\smallskip}
 & & Smith & Myers & Andr\'e & Smith & Myers & Andr\'e \\
\noalign{\smallskip}
\hline
\noalign{\smallskip}
\endhead
\noalign{\smallskip}
\hline
\noalign{\smallskip}
\multicolumn{8}{r}{Continued on next page} \\
\noalign{\smallskip}
\endfoot
\noalign{\smallskip}
\hline
\noalign{\smallskip}
\endlastfoot
\noalign{\smallskip}
 L\,1448\,NW                & 0   (yyy6) & 23 &  75    &  5 &  0.2 &    0.8 & 1.2    \\ 
 L\,1448\,C                 & 0   (yyy6) & 37 & 225    &  8 &  0.4 &      1 & 1.1    \\ 
 RNO\,15                    & 0   (yyy6) & 47 & 275    &160 &  0.6 &   $>$2 & 2.0    \\ 
 IRAS\,03256+3055           & 0$^\dagger$   (yyy5) & 36 & 100    & 50 & 0.05 &    0.4 & 0.4    \\ 
 NGC\,1333-I2               & 0   (yyy7) & 33 & 300    &  4 &    2 &   $>$2 & $>$3   \\ 
 NGC\,1333-I4\,A            & 0   (yyy6) & 29 & 200    &  6 &    1 &   $>$2 & $>$3   \\ 
 NGC\,1333-I4\,B            & 0   (yyy6) & 29 & 200    &  6 &    1 &   $>$2 & 3.0    \\ 
 IRAS\,03282+3035           & 0   (yyy5) & 37 & 110    &  8 & 0.06 &    0.4 & 0.7    \\ 
 HH\,211-MM                 & 0   (yyy6) & 25 &  75    &  8 &  0.3 &    0.8 & 0.8    \\ 
 IRAM\,04191                & 0   (yyy5) & 24 &  20    &  0 & 0.01 & $<$0.3 & 0.5    \\ 
 OMC\,3-MM\,6$^{*}$         & 0   (yyy7) & 48 & $>$500 &  8 &     5&   $>$2 & $>$3   \\ 
 HH\,212-MM                 & 0   (yyy5) & 35 & 175    & 60 &  0.4 &      1 & 0.45   \\ 
 HH\,25\,MMS                & 0   (yyy5) & 28 & 150    &  7 &  0.4 &      1 & 1.3    \\ 
 HH\,24\,MMS$^{*}$          & 0   (yyy5) & 53 & 300    &  6 &   0.8&   $>$2 & 3.0    \\ 
 NGC\,2068\,LBS\,17$^{*}$   & 0   (yyy5) & 27 &  75    & 0   & 0.10 & 0.4    &  2.5    \\
 NGC\,2264\,G-VLA\,2        & 0   (yyy5) & 44 & 250    & 25 &  0.6 &   $>$2 & 0.7    \\ 
 VLA\,1623$^{*}$            & 0   (yyy7) & 36 & 150    &  6 &   0.1&    0.5 & 0.9    \\ 
 IRAS\,16293-2422           & 0   (yyy7) & 29 & 200    &  5 &    1 &   $>$2 & $>$3   \\ 
 IRAS\,16544-1604           & 0   (yyy5) & 39 & 150    & 10 & 0.06 &    0.4 & 0.3    \\ 
 Serp-S68\,N                & 0   (yyy5) & 35 & 150    &  6 &  0.2 &    0.8 & 1.1    \\ 
 Serp-SMM\,4                & 0   (yyy6) & 38 & 200    &  6 &  0.2 &    0.8 & 1.5    \\ 
 Serp-SMM\,3                & 0   (yyy5) & 40 & 200    &  8 &  0.2 &    0.8 & 1.0    \\ 
 Serp-SMM\,2                & 0   (yyy5) & 47 & 200    & 40 &  0.2 &    0.7 & 0.4    \\ 
 IRAS\,19345+0727           & 0   (yyy6) & 29 & 125    &  8 &  0.2 &    0.7 & 0.8    \\ 
 IRAS\,20386+6751           & 0   (yyy5) & 29 & 150    &  8 &  0.5 &    1.1 & 1.2    \\ 
 L\,944-SMM\,1              & 0   (yyy5) & 53 & 200    & 18   & 0.21 & 0.7    &  0.5      \\
 Cep\,E-MM                  & 0   (yyy8) & 36 & 400    & 20 &    4 &   $>$2 & 2.8    \\ 
 IRAS\,23238+7401           & 0   (yyy5) & 36 & 100    & 20 & 0.04 &    0.4 & $<$0.2 \\ 
 \noalign{\smallskip}                                                                       
 \hline                                                                                     
 \noalign{\smallskip}                                                                       
 L\,1448-I2                 & 0/1 (yyn6) & 29 & 150    &  8 &  0.4 &    0.9 & 1.1    \\ 
 L\,1448\,N                 & 0/1 (yyn4) & 44 & 250    &  6 &  0.4 &      1 & 2.1    \\ 
 RNO\,15\,FIR               & 0/1 (yyn5) & 39 & 225    & 30 &  0.4 &      1 & 0.7    \\ 
 NGC\,1333\,I1              & 0/1 (?yy6) & 74 & 325    & 60 &    1 &   $>$2 & 0.8    \\ 
 B\,213                     & 0/1 (yyn4) & 41 &  90    & 5  & 0.01 & $<$0.3 &   0.2     \\ 
 IRAS\,04325+2402           & 0/1 (yyn4) & 43 & 150    & 30 & 0.04 &    0.4 & 0.7    \\ 
 L\,1527                    & 0/1 (yyn4) & 34 & 125    &  5 &  0.1 &    0.6 & 0.9    \\ 
 HH\,114\,MMS$^{*}$         & 0/1 (?yy6) & 70 & 350    & 10 &     1&   $>$2 & 1.6    \\ 
 IRAS\,05173-0555           & 0/1 (?yy5) &130 & 250    & 30 &  0.7 &    0.8 & 0.5    \\ 
 RNO\,43-MM                 & 0/1 (yyn5) & 35 & 200    & 20 &  0.3 &    0.9 & 0.6    \\ 
 L\,1641\,N                 & 0/1 (yyn5) & 43 & $>$500 &  9 &    4 &   $>$2 & $>$3   \\ 
 HH\,111\,MMS               & 0/1 (yyn5) & 54 & 300    & 20 &    1 &   $>$2 & 1.2    \\ 
 MonOB1\,IRAS12\,S1         & 0/1 (yyn5) & 20 & 150    &  8 &  7.8 &    1.3 & $>$3  \\
 IRAS\,15398-3359           & 0/1 (yyn4) & 36 & 100    &  7 & 0.04 &    0.4 & 0.4    \\ 
 IRAS\,18148-0440           & 0/1 (yyn5) & 34 & 200    & 20 &  0.5 &      1 & 0.8    \\ 
 Serp-FIRS\,1               & 0/1 (yyn6) & 32 & 300    &  8 &    2 &   $>$2 & $>$3   \\ 
 IRAS\,18331-0035           & 0/1 (yyn4) & 31 & 150    &  9 &  0.2 &    0.7 & 0.8    \\ 
 IRAS\,20050+2720           & 0/1 (yyn5) & 53 & $>$500 & 10 &   10 &   $>$2 & $>$3   \\ 
 IRAS\,21017+6742           & 0/1 (yyn5) & 43 & 125    & 40 & 0.04 &    0.4 & $<$0.2 \\ 
 CB\,230                    & 0/1 (yyn4) & 43 &  225   & 17 & 0.34 & 0.9    &  0.6   \\
 NGC\,7129\,FIRS\,2$^{*}$   & 0/1 (yyn4) & 49 &  $>$500 &  8 &   14 &  $>$2  &  $>$3  \\ 
 L\,1246-SMM\,1             & 0/1 (?yy5) & 290&  275   &  8 & 1.93 & 0.8    & 1.0    \\
 IRAS\,23385+6053$^{*}$     & 0/1 (yny7) & 36 & $>$500 &  0 & $>$50 & $>$2   & $>$3   \\
 \noalign{\smallskip}                                                                
 \hline                                                                              
 \noalign{\smallskip}                                                                
 L\,1551-IRS\,5             & 1   (nyn4) & 290& 350    &  9 &    6 &   $>$2 & 1.6    \\ 
 L\,1551-NE                 & 1   (nyn5) & 250& 225    & 10 &    1 &    0.7 & 0.7    \\ 
 OMC\,3-MM\,9               & 1   (nnn6) & 370& $>$500 & 20 &   50 &   $>$2 & 2.8    \\ 
 OMC\,3-MM\,7               & 1   (nyn4) & 380& $>$500 & 30 &   20 &   $>$2 & 1.7    \\ 
 HH\,147\,MMS               & 1   (nnn3) & 430& $>$500 &150 &   30 &   $>$2 & 2.0    \\ 
 IRAS\,08076-3556$^{*}$     & 1   (n?n5) & 365& 300    &200 &  4.2 &    1.0 & 2.5    \\
 Serp\,SMM\,5               & 1   (nyn4) & 350& 250    &  5 &    1 &    0.6 & 1.5    \\ 
 IRAS\,18148-0440           & 1   (nyn3) & 280& 250    &  8 &  0.9 &    0.6 & 0.8    \\ 
 L\,1251\,B                 & 1$^\dagger$   (nyn4) &  61& 250    &  7 &  0.4 &    1.0 & 2.0    \\
 L\,1211$^{*}$              & 1   (nyn4) & 426& $>$500 &  8 &   33 &   $>$2 & 1.2    \\
 \noalign{\smallskip}                                                                
\end{longtable}
\renewcommand{\arraystretch}{1.0}
\end{center}
\vspace{-0.75cm}
$^a$ Object name used in this paper. Further common names are given in
Table\,\ref{sources}. The $^{*}$ markes objects where the SED is is only well
determined on one side of the peak and for the other side only upper limits or
very few data is available. \\ 
$^b$ Classification due to the procedure described in
Sect.\,\ref{classification}. In brackets we give as 'y', 'n', or '?' the
answers for our three different tests if the object is a Class\,0 source. The
ending number gives the orders of magnitude between the maximum of the emission
and the NIR detection/upper limit. \\ 
$^c$ Ages and final masses obtained from the evolutionary models of Smith
\cite{1998Ap&SS.261..169S, 2000IrAJ...27...25S, s02}, Myers et al.
\cite{1998ApJ...492..703M}, and Andr\'e et al. \cite{2000prpl.conf...59A}. \\
$^\dagger$ Classification uncertain, since no or no conclusive sub-mm or 
millimeter maps are available. \\
\twocolumn


\begin{appendix}

\onecolumn
\section{Source Sample}
\label{sourcesample}
\renewcommand{\arraystretch}{0.8}
\begin{center}
\begin{longtable}{llllcrc}
\caption{\label{sources} \bfseries Complete source sample} \\
\noalign{\smallskip}
\hline
\noalign{\smallskip}
Object$^a$ & Ref.\,Obj.$^b$ & RA\,(2000) & Dec\,(2000) & Ref.$^c$ & d\,[pc] &
Ref.$^c$ \\ 
\noalign{\smallskip}
\endfirsthead
\multicolumn{7}{c}{\bfseries \tablename\ \thetable{} -- continued from previous
page} \\ 
\noalign{\smallskip}
\hline
\noalign{\smallskip}
Object$^a$ & Ref.\,Obj.$^b$ & RA\,(2000) & Dec\,(2000) & Ref.$^c$ & d\,[pc] &
Ref.$^c$ \\ 
\noalign{\smallskip}
\hline
\noalign{\smallskip}
\endhead
\noalign{\smallskip}
\hline
\noalign{\smallskip}
\multicolumn{7}{r}{Continued on next page} \\
\noalign{\smallskip}
\endfoot
\hline
\endlastfoot
\hline
\noalign{\smallskip}	
W3\,OH              & {15}, {43}                                   	           & 02 27 04.72 & $+$61 52 24.7 & {71} & 2200   & {43} \\
\multicolumn{7}{l}{\hspace{0.5cm}\small TW\,3\,(H2O)} \\
L\,1448-I2          & {22}, {43}                             	           & 03 25 22.5  & $+$30 45 06   & {59} & 300    & {54} \\
\multicolumn{7}{l}{\hspace{0.5cm}\small IRAS\,03222+3034} \\
L\,1448\,NW         & {48}, {54}, {70}                                     & 03 25 35.65 & $+$30 45 34.2 & {47} & 300    & {54} \\
\multicolumn{7}{l}{\hspace{0.5cm}\small LDN\,1448\,IRS\,3C} \\
L\,1448\,N          & {16}, {22}, {40}, {43}, {48}, {54}                    & 03 25 36.3  & $+$30 45 16   & {48} & 300    & {54} \\
\multicolumn{7}{l}{\hspace{0.5cm}\small LDN\,1448\,IRS\,3, IRAS\,03225+3034} \\
L\,1448\,C          & {16}, {22}, {40}, {43}, {48}, {54}, {70}              & 03 25 38.8  & $+$30 44 00   & {48} & 300    & {54} \\
\multicolumn{7}{l}{\hspace{0.5cm}\small L\,1448-MM} \\
RNO\,15\,FIR        & {48}, {64}, {70}                                     & 03 27 39.0  & $+$30 12 59   & {71} & 350    & {70} \\
\multicolumn{7}{l}{\hspace{0.5cm}\small LDN\,1455\,FIR} \\
RNO\,15             & {70} {\small(?)}                         	           & 03 27 42.9  & $+$30 12 27   & {70} & 350    & {70} \\
\multicolumn{7}{l}{\hspace{0.5cm}\small L\,1455, B\,204} \\
NGC\,1333\,I1       & {70} {\small(Class0/1)}                  	           & 03 28 38.7  & $+$31 13 32   & {71} & 350    & {54} \\
\multicolumn{7}{l}{\hspace{0.5cm}\small IRAS\,03255+3103} \\
IRAS\,03256+3055    & {68} {\small(Class1)}                    	           & 03 28 44.5  & $+$31 05 39.7 & {68} & 400    & { 2} \\
NGC\,1333-I2        & {40}, {43}, {54}, {70}                               & 03 28 55.4  & $+$31 14 35   & {58} & 350    & {54} \\
\multicolumn{7}{l}{\hspace{0.5cm}\small IRAS\,03258+3104} \\
SVS\,13\,B          & {22} {\small(OF)}, {43}, {48} {\small(Class1)}, {49}   & 03 29 03.06 & $+$31 15 51.7 & {47} & 350    & {70} \\
NGC\,1333-I4\,A     & {22}, {40}, {43}, {54}                               & 03 29 12.04 & $+$31 13 30.5 & {70} & 350    & {54} \\
NGC\,1333-I4\,B     & {22}, {43}, {54}                                     & 03 29 13.6  & $+$31 13 06.6 & {70} & 350    & {54} \\
IRAS\,03282+3035    & {16}, {43}, {48}, {54}                               & 03 31 20.3  & $+$30 45 25   & {48} & 300    & {54} \\
IC\,348\,MMS        & {63}                                    	           & 03 43 56.9  & $+$32 03 06   & {63} & 300    & {54} \\
HH\,211-MM          & {43}, {54}, {64}, {70}                               & 03 43 56    & $+$32 00 48   & {54} & 300    & {54} \\
B\,213              & {16} {\small(OF)}, {48} {\small(Class1)},  & 04 19 43.00 & $+$27 13 33.7 & {60} & 140    & {54} \\
\multicolumn{1}{l}{\hspace{0.5cm}\small IRAS\,04166+2706} & \multicolumn{6}{l}{{54} {\small(0/I)}, {68}}\\
IRAM\,04191         & {43}, {54}                                           & 04 21 56.9  & $+$15 29 46   & {36} & 140    & {54} \\
L\,1551-IRS\,5      & {13}, {22} {\small(OF)}                               & 04 31 34.15 & $+$18 08 05.2 & {44} & 140    & {54} \\
\multicolumn{7}{l}{\hspace{0.5cm}\small IRAS\,04287+1801} \\
L\,1551-NE          & {13}, {22}                                           & 04 31 44.44 & $+$18 08 32   & {37} & 140    & {54} \\
IRAS\,04325+2402    & {26} {\small(T$_{\rm bol}$\,=\,157\,K)} 	            & 04 35 35.0  & $+$24 08 22   & {54} & 140    & { 2} \\
\multicolumn{7}{l}{\hspace{0.5cm}\small LDN\,1535\,IRS} \\
L\,1527            & {13}, {16}, {43}, {48}, {49}, {54}                    & 04 39 53.9  & $+$26 03 11   & {48} & 140    & {54} \\
\multicolumn{7}{l}{\hspace{0.5cm}\small IRAS\,04368+2557} \\
HH\,114\,MMS        & {24}, {43}                          	           & 05 18 15.21 & $+$07 12 03   & {21} & 450    & {43} \\
IRAS\,05173-0555    & {70}                                     	           & 05 19 48.9  & $-$05 52 05   & {61} & 460    & {70} \\
\multicolumn{7}{l}{\hspace{0.5cm}\small RNO\,40\,FIR, L\,1634} \\
L\,1634\,IRS\,7     & {70} {\small(Class0?)}                   	            & 05 19 51.5  & $-$05 52 06   & {70} & 460    & {70} \\
RNO\,43-MM          & { 7}, {43}                                     	           & 05 32 19.4  & $+$12 49 41   & {32} & 400    & {43} \\
\multicolumn{7}{l}{\hspace{0.5cm}\small IRAS\,05295+1247} \\
OMC\,3-MM\,6        & {25}, {43}                                           & 05 35 23.4  & $-$05 01 29   & {41} & 450    & {43} \\
OMC\,3-MM\,9        & {25}                                    	           & 05 35 25.9  & $-$05 05 44   & {41} & 450    & {43} \\
OMC\,3-MM\,7        & {25}                                    	           & 05 35 26.5  & $-$05 03 55   & {41} & 450    & {43} \\
\multicolumn{7}{l}{\hspace{0.5cm}\small IRAS\,05329-0505} \\
HH\,1-2\,MMS\,3     & {49}                                    	           & 05 36 18.2  & $-$06 45 45.3 & {49} & 460    & {49} \\
L\,1641\,N          & {22} {\small(OF)}, {70}                               & 05 36 18.6  & $-$06 22 10   & {70} & 390    & {70} \\
\multicolumn{7}{l}{\hspace{0.5cm}\small IRAS\,05338-0624} \\
HH\,1-2\,MMS\,2     & {49}                                    	           & 05 36 18.8  & $-$06 45 25.3 & {49} & 460    & {49} \\
L\,1641-VLA\,1      & {43}, {49}                                           & 05 36 22.8  & $-$06 46 07.6 & {49} & 460    & {49} \\
\multicolumn{7}{l}{\hspace{0.5cm}\small HH\,1-2\,MMS\,1} \\
L\,1641\,SMS\,III   & {70} {\small(Class0/1)}                  	            & 05 36 24.0  & $-$06 24 54   & {70} & 390    & {70} \\
HH\,147\,MMS        & {49}                                    	           & 05 36 25.2  & $-$06 44 39.8 & {49} & 460    & {49} \\
\multicolumn{7}{l}{\hspace{0.5cm}\small IRAS\,05339-0646} \\
L\,1641\,SMS\,IV    & {70} {\small(Class0?)}                   	            & 05 36 41.8  & $-$06 26 12   & {70} & 390    & {70} \\
NGC\,2023-MM\,1     & {19}, {43}                                     	           & 05 41 24.54 & $-$02 18 09   & {59} & 450    & {43} \\
NGC\,2024-FIR\,5    & { 6}, {43}                                     	           & 05 41 44.22 & $-$01 55 41.32& {65} & 450    & {43} \\
NGC\,2024-FIR\,6    & { 6}, {43}                                     	           & 05 41 45.17 & $-$01 56 00.56& {65} & 450    & {43} \\
HH\,212-MM          & { 7}, {43}                                     	           & 05 43 51.1  & $-$01 03 01   & {71} & 400    & {43} \\
\multicolumn{7}{l}{\hspace{0.5cm}\small IRAS\,05413-0104} \\
HH\,25\,MMS         & {12}, {43}                                     	           & 05 46 07.8  & $-$00 13 41   & {52} & 450    & {43} \\
HH\,24\,MMS         & {22}, {43}                                           & 05 46 08.8  & $-$00 10 47   & {52} & 450    & {43} \\
NGC\,2068\,LBS\,17  & {45}                                                 & 05 46 30.8 &  $-$00 02 40   & {56} & 400    & {45} \\
HH\,111\,MMS        & {49}                                    	           & 05 51 46.3  & $+$02 48 28   & {49} & 460    & {49} \\
\multicolumn{7}{l}{\hspace{0.5cm}\small IRAS\,05491+0247, LDN\,1617\,1} \\
MonOB1\,IRAS12\,S1  & {69}                                                 & 06 41 05.8  & $+$09 34 09   & {69} & 800    & {69} \\
NGC\,2264\,G-VLA\,2 & {14}, {43}                                     	           & 06 41 10.9  & $+$09 56 02   & {14} & 800    & {43} \\
\multicolumn{7}{l}{\hspace{0.5cm}\small IRAS\,06384+0958} \\
IRAS\,08076-3556    & {11}, {43}                                     	           & 08 09 32.8  & $-$36 05 00   & {71} & 400    & {43} \\
\multicolumn{7}{l}{\hspace{0.5cm}\small BHR\,12} \\
Ced\,110\,IRS\,10   & {53}                                    	           & 11 06 32    & $-$77 23 42   & {20} & 150    & {53} \\
\multicolumn{7}{l}{\hspace{0.5cm}\small Cha-MMS\,1} \\
Ced\,110\,IRS\,4    & {26}, {53}                                           & 11 06 47.14 & $-$77 22 34.0 & {55} & 150    & {53} \\
\multicolumn{7}{l}{\hspace{0.5cm}\small IRAS\,11051-7706} \\
BHR\,71-MM          & {23}, {43}                                     	           & 12 01 37    & $-$65 08 54   & {71} & 200    & {43} \\
\multicolumn{7}{l}{\hspace{0.5cm}\small IRAS\,11590-6452} \\
IRAS\,12500-7658    & {26}                                    	           & 12 53 38.9  & $-$77 15 53   & {57} & 180    & { 6} \\
IRAS\,12533-7632    & {26}                                    	           & 12 57 00.1  & $-$76 48 35   & {52} & 180    & { 6} \\
IRAS\,12553-7651    & {26}                                    	           & 12 59 05.5  & $-$77 07 34   & {71} & 180    & {30} \\
IRAS\,12554-7635    & {26}                                    	           & 12 59 13.4  & $-$76 51 11   & {71} & 180    & { 6} \\
IRAS\,13036-7644    & {26}, {43}                                           & 13 07 36.0  & $-$77 00 04.4 & {29} & 200    & {43} \\
\multicolumn{7}{l}{\hspace{0.5cm}\small BHR\,86} \\
IRAS\,15398-3359    & {26}, {48}                                           & 15 43 01.3  & $-$34 09 12   & {29} & 130    & {48} \\
\multicolumn{7}{l}{\hspace{0.5cm}\small B\,228} \\
IRAS\,16017-3936    & {26}                                    	           & 16 05 04.7  & $-$39 45 04   & {71} & 130    & {26} \\
VLA\,1623           & {16}, {22}, {43}                                     & 16 26 26.32 & $-$24 24 30.1 & {47} & 160    & {43} \\
IRAS\,16293-2422    & {13}, {16}, {22}, {43}                               & 16 32 22.8  & $-$24 28 33   & {71} & 160    & {43} \\
\multicolumn{7}{l}{\hspace{0.5cm}\small L\,1689} \\
IRAS\,16493-4242    & {26}                                    	           & 16 52 56.6  & $-$42 47 56   & {71} & 130    & {26} \\
IRAS\,16544-1604    & {39}                                    	           & 16 57 19.5  & $-$16 09 21   & {39} & 160    & {28} \\
\multicolumn{7}{l}{\hspace{0.5cm}\small L\,146, CB\,68} \\
Trifid-TC\,3        & {31}, {43}                                     	           & 18 02 07    & $-$23 05 11   & {71} & 1680   & {43} \\
IRAS\,18148-0440    & {40}, {43}, {48}, {54}, {62}                          & 18 17 29.8  & $-$04 39 38   & {71} & 200    & {54} \\
\multicolumn{7}{l}{\hspace{0.5cm}\small L\,483} \\
Serp-S68\,N         & {18}, {40}, {43}                                     & 18 29 48.1  & $+$01 16 41   & {38} & 310    & {43} \\
Serp-FIRS\,1        & {18}, {40}, {43}                                     & 18 29 49.79 & $+$01 15 18.6 & {38} & 310    & {43} \\
\multicolumn{7}{l}{\hspace{0.5cm}\small IRAS\,18273+0113} \\
Serp\,SMM\,5        & {40}                                     	           & 18 29 51.1  & $+$01 16 36   & {38} & 310    & {43} \\
Serp-SMM\,4         & {18}, {40}, {43}                                     & 18 29 56.5  & $+$01 13 10   & {38} & 310    & {43} \\
Serp-SMM\,3         & {18}, {40}, {43}                                     & 18 29 59.2  & $+$01 13 58   & {38} & 310    & {43} \\
Serp-SMM\,2         & {18}, {40}                                           & 18 30 00.2  & $+$01 12 57   & {38} & 310    & {43} \\
IRAS\,18331-0035    & {40}, {49}                                           & 18 35 42.0  & $-$00 33 18   & {71} & 310    & {24} \\
\multicolumn{7}{l}{\hspace{0.5cm}\small HH\,108/109\,IRS, L\,558} \\
HH\,108\,MMS        & {49}                                    	           & 18 35 46.55 & $-$00 32 41.8 & {71} & 310    & {24} \\
G\,34.24+0.13\,MM   & {34}, {43}                                     	           & 18 53 21.4  & $+$01 13 45.3 & {34} & 3700   & {43} \\
R\,CrA\,R1          & {22}                                     	           & 19 01 55.8  & $-$36 57 29.6 & {22} & 130    & {22} \\
IRAS\,19156+1906    & {16}, {40}, {43}, {48}, {54}                         & 19 17 53.16  & $+$19 12 16.6 & {71} & 300    & {54} \\
\multicolumn{7}{l}{\hspace{0.5cm}\small L\,723} \\
IRAS\,19180+1114    & {40}, {62}                               	           & 19 20 25.8  & $+$11 19 52   & {71} & 300    & { 3} \\
\multicolumn{7}{l}{\hspace{0.5cm}\small L\,673\,A, RNO\,109} \\
IRAS\,19345+0727    & {16}, {22}, {40}, {43}, {48}, {54}, {62}              & 19 37 01.03  & $+$07 34 10.9 & {71} & 250    & {54} \\
\multicolumn{7}{l}{\hspace{0.5cm}\small B\,335, LDN\,663} \\
IRAS\,20050+2720    & {40}, {49}                                           & 20 07 06.1  & $+$27 28 59   & {49} & 700    & {49} \\
S\,106-SMM          & { 9}, {43}                                     	           & 20 27 11.8  & $+$37 22 45   & {17} & 600    & {43} \\
IRAS\,20386+6751    & {40}, {43}, {48}, {49}, {54}, {64}                   & 20 39 06.5   & $+$68 02 13   & {48} & 440    & {54} \\
\multicolumn{7}{l}{\hspace{0.5cm}\small L\,1157} \\
GF\,9-2             & {10}, {43}                                     	           & 20 51 30.1  & $+$60 18 39   & {71} & 200    & {43} \\
IRAS\,21017+6742    & {48}, {62}                               	           & 21 02 23.1  & $+$67 54 19   & {48} & 288    & {48} \\
\multicolumn{7}{l}{\hspace{0.5cm}\small L\,1172} \\
CB\,230             & {68}                                                 & 21 17 40.0   & $+$68 17 32   & {68} & 450   & {68} \\
\multicolumn{7}{l}{\hspace{0.5cm}\small IRAS\,21169+6804, LDN\,1177} \\
L\,944-SMM\,1       & {62}                                                & 21 17 40.7   & $+$43 18 08.5   & {62} & 700    & { 1} \\
IC\,1396\,W         & {64}                                     	           & 21 26 06.1  & $+$57 56 17   & {71} & 750    & {64} \\
\multicolumn{7}{l}{\hspace{0.5cm}\small IRAS\,21246+5743} \\
CB\,232             & {39}                                    	           & 21 37 10.8  & $+$43 20 39.4 & {39} & 350    & {27} \\
\multicolumn{7}{l}{\hspace{0.5cm}\small IRAS\,21352+4307} \\
NGC\,7129\,FIRS\,2  & {33}, {40}                                           & 21 43 01.6   & $+$66 03 26   & {50} & 1250   & {33} \\
L\,1251\,B          & {68}                                                 & 22 38 42.52  & $+$75 11 45.6 & { 8} &  300   & { 8} \\
\multicolumn{7}{l}{\hspace{0.5cm}\small IRAS\,22376+7455} \\
L\,1211             & {64}                                     	           & 22 47 17.2  & $+$62 02 34   & {42} & 725    & {64} \\
\multicolumn{7}{l}{\hspace{0.5cm}\small IRAS\,22451+6154} \\
Cep\,E-MM           & {43}, {49}, {64}                                     & 23 03 13.1  & $+$61 42 26   & {51} & 730    & {43} \\
\multicolumn{7}{l}{\hspace{0.5cm}\small IRAS\,23011+6126} \\
NGC\,7538\,S        & {66}                                    	           & 23 13 44.98 & $+$61 26 49.2 & {66} & 3500   & {46} \\
L\,1246-SMM\,1      & {62}                                                 & 23 25 04.7  & $+$63 36 40   & {62} & 730    & {62} \\
IRAS\,23238+7401    & {48}, {62}                               	           & 23 25 46.4  & $+$74 17 38   & {48} & 180    & {48} \\
\multicolumn{7}{l}{\hspace{0.5cm}\small CB\,244, LDN\,1262\,4} \\
IRAS\,23385+6053    & {35}, {43}                                     	           & 23 40 54.5  & $+$61 10 28   & {67} & 4900   & {43} \\
\noalign{\smallskip}
\end{longtable}
\end{center}
\vspace{-1.1cm}
$^a$ Object name used in this paper. Further common names are given below. \\ 
$^b$ References where the object was mentioned as candidate or confirmed
Class\,0 source (in the samples listed in Sect.\,\ref{sample}). In brackets
deviations from the general selection criteria are shown. In case the review
article from Andr\'e et al. \cite{2000prpl.conf...59A} provides the only
reference, we additionally list the discovery paper of the source. \\ 
$^c$ References for the origin of the position (Col.\,5) and distance (Col.\,7).\\ 
{\bfseries References:} 
{\bf (1) } Dame \& Thaddeus           \cite{1985ApJ...297..751D},
{\bf (2) } Clark                      \cite{1991ApJS...75..611C}, 
{\bf (3) } Ladd et al.                \cite{1991ApJ...366..203L}, 
{\bf (4) } Carballo et al.            \cite{1992A&A...262..106C},
{\bf (5) } Mezger et al.              \cite{1992A&A...256..631M},
{\bf (6) } Prusti et al.              \cite{1992A&A...260..151P},
{\bf (7) } Zinnecker et al.           \cite{1992A&A...265..726Z},
{\bf (8) } Kun \& Prusti              \cite{1993A&A...272..235K},
{\bf (9) } Richer et al.              \cite{1993MNRAS.262..839R},
{\bf (10)} G\"usten                   \cite{1994coun.conf..169G},
{\bf (11)} Persi et al.               \cite{1994A&A...282..233P},
{\bf (12)} Bontemps et al.            \cite{1995A&A...297...98B},
{\bf (13)} Chen et al.                \cite{1995ApJ...445..377C}, 
{\bf (14)} Ward-Thompson et al.       \cite{1995MNRAS.273L..25W}, 
{\bf (15)} Wilner et al.              \cite{1995ApJ...449L..73W},
{\bf (16)} Bontemps et al.            \cite{1996A&A...311..858B}, 
{\bf (17)} Holland et al.             \cite{1996A&A...309..267H}, 
{\bf (18)} Hurt et al.                \cite{1996ApJ...460L..45H},
{\bf (19)} Launhardt et al.           \cite{1996A&A...312..569L}, 
{\bf (20)} Reipurth et al.            \cite{1996A&A...314..258R},
{\bf (21)} Rodr\'{i}guez \& Reipurth \cite{1996RMxAA..32...27R}, 
{\bf (22)} Saraceno et al.            \cite{1996A&A...309..827S},
{\bf (23)} Bourke et al.              \cite{1997ApJ...476..781B}, 
{\bf (24)} Chini et al.               \cite{1997A&A...325..542C}, 
{\bf (25)} Chini et al.               \cite{1997ApJ...474L.135C}, 
{\bf (26)} Chen et al.                \cite{1997ApJ...478..295C}, 
{\bf (27)} Codella \& Muders          \cite{1997MNRAS.291..337C}, 
{\bf (28)} Launhardt \& Henning       \cite{1997A&A...326..329L}, 
{\bf (29)} Mardones et al.            \cite{1997ApJ...489..719M},
{\bf (30)} Olmi et al.                \cite{1997A&A...326..373O}, 
{\bf (31)} Cernicharo et al.          \cite{1998Sci...282..462C},
{\bf (32)} Dent et al.                \cite{1998MNRAS.301.1049D}, 
{\bf (33)} Eiroa et al.               \cite{1998A&A...335..243E}, 
{\bf (34)} Hunter et al.              \cite{1998ApJ...493L..97H},
{\bf (35)} Molinari et al.            \cite{1998ApJ...505L..39M}, 
{\bf (36)} Andr\'e et al.             \cite{1999ApJ...513L..57A}, 
{\bf (37)} Devine et al.              \cite{1999AJ....118..972D},
{\bf (38)} Davis et al.               \cite{1999MNRAS.309..141D},
{\bf (39)} Huard et al.               \cite{1999ApJ...526..833H}, 
{\bf (40)} Park et al.                \cite{1999ApJ...520..223P}, 
{\bf (41)} Reipurth et al.            \cite{1999AJ....118..983R},
{\bf (42)} Tafalla et al.             \cite{1999A&A...348..479T},
{\bf (43)} Andr\'e et al.             \cite{2000prpl.conf...59A}, 
{\bf (44)} Chandler \& Richer         \cite{2000ApJ...530..851C}, 
{\bf (45)} Gibb \& Little             \cite{2000MNRAS.313..663G},
{\bf (46)} Kalenskii et al.           \cite{2000A&A...354.1036K},
{\bf (47)} Looney et al.              \cite{2000ApJ...529..477L}, 
{\bf (48)} Shirley et al.             \cite{2000ApJS..131..249S}, 
{\bf (49)} Chini et al.               \cite{2001A&A...369..155C}, 
{\bf (50)} Fuente et al.              \cite{2001A&A...366..873F}, 
{\bf (51)} Giannini et al.            \cite{2001ApJ...555...40G}, 
{\bf (52)} Johnstone et al.           \cite{2001ApJ...559..307J}, 
{\bf (53)} Lehtinen et al.            \cite{2001A&A...367..311L}, 
{\bf (54)} Motte \& Andr\'e           \cite{2001A&A...365..440M}, 
{\bf (55)} Persi et al.               \cite{2001A&A...376..907P}, 
{\bf (56)} Phillips et al.            \cite{2001MNRAS.326..927P},
{\bf (57)} Voung et al.               \cite{2001A&A...379..208V}, 
{\bf (58)} Ward-Thompson \& Buckley   \cite{2001MNRAS.327..955W}, 
{\bf (59)} Anglada \& Rodr\'{i}guez   \cite{2002RMxAA..38...13A}, 
{\bf (60)} Hartmann                   \cite{2002ApJ...578..914H}, 
{\bf (61)} Nisini et al.              \cite{2002A&A...393.1035N}, 
{\bf (62)} Visser et al.              \cite{2002AJ....124.2756V},
{\bf (63)} Eisl\"offel et al.         \cite{2003ApJ...595..259E}, 
{\bf (64)} Froebrich et al.           \cite{2003MNRAS.346..163F}, 
{\bf (65)} Rodr\'{i}guez et al.       \cite{2003ApJ...598.1100R},  
{\bf (66)} Sandell et al.             \cite{2003ApJ...590L..45S},
{\bf (67)} Thompson \& Macdonald      \cite{2003A&A...407..237T},
{\bf (68)} Young et al.               \cite{2003ApJS..145..111Y}, 
{\bf (69)} Wolf-Chase et al.          \cite{2003MNRAS.344..809W},
{\bf (70)} Rengel et al.              \cite{2004A&A..inprep.R}, 
{\bf (71)} SIMBAD database
\renewcommand{\arraystretch}{1.0} \twocolumn

\onecolumn
\section{Notes to individual objects}
\label{notesonsources}
\renewcommand{\arraystretch}{1.0}
\begin{center}
\begin{longtable}{lp{12cm}}
\caption{\label{notes} \bfseries Notes to individual sources} \\ 
\noalign{\smallskip}
\hline
\noalign{\smallskip}
Object$^a$ & Description \\ 
\noalign{\smallskip}
\hline
\noalign{\smallskip}
\endfirsthead
\noalign{\smallskip}
\multicolumn{2}{c}{\bfseries \tablename\ \thetable{} -- continued from previous
page} \\ 
\noalign{\smallskip}
\hline
\noalign{\smallskip}
Object$^a$ & Description \\ 
\noalign{\smallskip}
\hline
\noalign{\smallskip}
\endhead
\noalign{\smallskip}
\hline
\noalign{\smallskip}
\multicolumn{2}{r}{Continued on next page} \\
\noalign{\smallskip}
\endfoot
\noalign{\smallskip}
\hline
\noalign{\smallskip}
\endlastfoot
\noalign{\smallskip}

W3\,OH & The 2MASS source is 3.5\arcsec\, away from the coordinates adopted
here. \\

L1448-I2 & Fluxes from O'Linger et al. \cite{1999ApJ...515..696O} taken in the
whole observed field are not used. \\

L\,1448\,NW & L\,1448\,NW is situated about 20\arcsec\, NNW of the binary
L\,1448\,N(A/B). The flux measurements are strongly influenced by the extended
envelope of this nearby source. Fluxes from Chandler \& Richer
\cite{2000ApJ...530..851C} are given in Jy/beam and hence not used. The two
2.7\,mm fluxes measured in small ($<$\,5\arcsec) apertures by Looney et al.
\cite{2000ApJ...529..477L} are not used either.  \\

L\,1448\,N & This source is actually a protostellar binary consisting of
L\,1448 IRAS\,3A and IRAS\,3B. Since IRAS\,3A is about one order of magnitude
brighter in the FIR as well as in the millimeter range, all the unresolved
observations are attributed to IRAS\,3A. Sub-mm fluxes taken in 40\arcsec\,
apertures (Shirley et al. \cite{2000ApJS..131..249S}) are not used due to the
vicinity of L\,1448\,NW. \\

L\,1448\,C & We chose the values measured in boxes larger than 30\arcsec, in
order not to miss emission from the outer regions of the extended envelope. \\

RNO\,15\,FIR & In the vicinity (about 1\arcmin\, south-east) of RNO\,15\,FIR
the somewhat older source RNO\,15 can be found. Both objects possess about the
same brightness and hence flux measurements in large boxes might be influenced
by RNO\,15. This applies especially for the IRAS data. At 60 and 100\,$\mu$m
there are discrepancies between the IRAS and ISO data. We used the ISO data
since they are measured in a slightly smaller aperture. \\

RNO\,15 & The data at short wavelengths and arround the maximum show a large
scatter. This might partly be due to the nearby source RNO\,15\,FIR and also
due to the fact that most of these measurements are taken before 1990 and
suffer from large uncertainties. \\

NGC\,1333\,I1 & Here the fluxes given by Sandell et al.
\cite{2001ApJ...546L..49S} are excluded. They are unreliable because the source
is at the edge of their observed field. \\

IRAS\,03256+3055 & This source seems to consist of three objects (Young et al.
\cite{2003ApJS..145..111Y}). The IRAS position coincides with the weakest of
the sub-mm sources. We excluded the fluxes measured in the 120\arcsec\,
apertures. There is a discrepancy between the 800 and 850\,$\mu$m point. We did
not use the 800\,$\mu$m data from Anglada \& Rodr\'{i}guez
\cite{2002RMxAA..38...13A} since an unusual $\beta$\,$>$\,3 would be obtained.
\\

NGC\,1333\,I2 & NGC\,1333\,I2 is a double source consisting of IRAS\,2A and 2B
with a separation of about 30\arcsec. IRAS\,2A is much brighter than IRAS\,2B
in the sub-mm and hence the fluxes given for both objects are attributed to
IRAS\,2A. \\

SVS\,13\,B & Here it is particular important to have the aperture sizes in
which the photometry was obtained, since SVS\,13\,B is a triple protostellar
system. It consists of MMS\,1, 2, and 3. While MMS\,1 and 2 have about the same
brightness, MMS\,3 is weaker. There are no data for the single sources at
wavelengths shorter than 350\,$\mu$m. \\

NGC\,1333\,I4\,A & This is a double source I4\,A1 and I4\,A2. I4\,A1 is the
brighter of the two sources and all the photometric data was attributed to this
object. \\

NGC\,1333\,I4\,B & NGC\,1333\,I4\,B is situated about 30\arcsec\, south-east of
NGC\,1333\,I4\,A and hence the fluxes measured in large areas might be
influenced. The data from Sandell et al. \cite{2001ApJ...546L..49S} were taken
in a relatively small aperture and might not include all the emission from the
envelope. It was therefore not used. The flux from Rengel et al.
{\cite{2004A&A..inprep.R}} is also excluded since the object is at the border of
their map. The large fluxes given from Rebull et al. \cite{2003AJ....125.2568R}
are not used since they might be influenced by surrounding emission. \\

IRAS\,03282+3035 & Even though this source seems to be isolated, the fluxes
measured in the 120\arcsec\, box by Sandell et al. \cite{2000ApJS..131..249S}
are excluded. \\

HH\,211\,MM & The data from Dent et al. \cite{1998MNRAS.301.1049D} are taken in
an unknown aperture and hence not used. \\

B\,213 & We did not use the fluxes measured in a 120\arcsec\, aperture by Young
et al. \cite{2003ApJS..145..111Y} since they might by influenced by other
surrounding sources. The IRAS 100\,$\mu$m limit was used as flux. \\

L\,1551-IRS\,5 & Mesurements taken in very small apertures ($<$\,10\arcsec)
were exluded. Also the data taken in the 270\arcsec\, apertures are not used.
\\

L\,1551-NE & This object is a double source consisting of L\,1551-NE\,A and B.
All fluxes are attributed to L\,1551-NE\,A. \\

IRAS\,04325+2402 & The measurements arround 800\,$\mu$m and 1.3\,mm do not
correspond to each other. We did not use the low values given by
Moriarty-Schieven et al. \cite{1994ApJ...436..800M}, since very high values of
$\beta$ ($>$\,3) are needed to explain these data. Also the upper limits given
by Barsony \& Kenyon \cite{1992ApJ...384L..53B} seem to be too low in this
context. The 2MASS source is 6\arcsec\, away from the coordinates adopted here.
\\

L\,1527 & There is a large scatter in the fluxes taken between
350\,$\mu$m and 1.3\,mm even if they are taken with the same aperture size. We
do not use data taken in the 120\arcsec\, apertures. \\

HH\,114\,MMS & Difficulties occur since the IRAS points are only upper limits
at all wavelengths. The 1.3\,mm flux given by Chini et al.
{\cite{1997A&A...325..542C}} for the whole observed field was exluded. \\

IRAS\,05173-0555 & There is a large scatter in the data at all wavelengths. We
did not use the flux of the whole field given by Chini et al.
{\cite{1997A&A...325..542C}}. There are two sets of fluxes in the sub-mm data;
low and high values. We chose the lower fluxes since they seem to present the
better the central source and its envelope. \\

L\,1634\,IRS\,7 & The 2MASS source is 3\arcsec\, away from the coordinates
adopted here. \\

RNO\,43-MM & The measurement from Chini et al. {\cite{1997A&A...325..542C}} at
1.3\,mm for the whole field was excluded. \\

OMC\,3-MM6 & No data shortward of 100\,$\mu$m are available. The 185\,$\mu$m
point taken in a 3\arcmin\, aperture was not used. All the fluxes below
200\,$\mu$m seem to be anomolous due to the large apertures. \\

OMC\,3-MM9 & There is no IRAS detection. The 2MASS source is 6.6\arcsec\, away
from the coordinates adopted here. \\

OMC\,3-MM7 & The upper limit at 100\,$\mu$m in the IRAS catalogue seems to be
too small compared to the other data. The 2MASS source is 3.5\arcsec\, away
from the coordinates adopted here. \\

L\,1641\,N & The 2MASS source is 7\arcsec\, away from the coordinates adopted
here. \\

HH\,1-2\,MMS\,2 & There is a large scatter in the 1.3\,mm data. We used only
the flux measured in the 8.5\arcsec\, aperture. \\

L\,1641\,VLA\,1 & The scatter in the measurements is very large. We did not use
the huge 1.3\,mm flux. The 2MASS source is 4.3\arcsec\, away from the
coordinates adopted here. \\

L\,1641\,SMS\,III & The 2MASS source is 6.5\arcsec\, away from the coordinates
adopted here. \\

HH\,147\,MMS & Measurements taken in large apertures ($>$\,30\arcsec) are
excluded. The 2MASS source is 2.5\arcsec\, away from the coordinates adopted
here. \\

NGC\,2023-MM\,1 & The 450\,$\mu$m point from Sandell et al.
\cite{1999ApJ...519..236S} in the small aperture was excluded, as well as the
data taken in the 180\arcsec\, apertures. \\

NGC\,2024-FIR\,5 & The huge 1.3\,mm flux in the 30\arcsec\, aperture was
excluded. \\

HH\,212-MM & The 1.3\,mm flux from Chini et al. {\cite{1997A&A...325..542C}},
measured in the whole observed field, was excluded.  \\

HH\,25\,MMS & The two fluxes from Gibb \& Davis \cite{1998MNRAS.298..644G}
measured in large apertures at 60 and 100\,$\mu$m are excluded. The 1.3\,mm
point is not used. \\

HH\,24\,MMS & All the IRAS points are just upper limits. The 1.3\,mm flux from
Launhard et al. {\cite{1996A&A...312..569L}} measured in a 30\arcsec\, aperture
was excluded. \\

NGC\,2068\,LBS\,17 & All data shortward of 100\,$\mu$m are upper limits. \\

HH\,111\,MMS & The data obtained in large apertures ($\ge$\,30\arcsec) were
excluded as well as the high flux at 1.3\,mm. The 2MASS source is 4\arcsec\,
away from the coordinates adopted here. \\

MonOB1\,IRAS12\,S1 & The 2MASS source is 4\arcsec\, away from the coordinates
adopted here. \\

IRAS\,08076-3556 & The 2MASS source is 4.7\arcsec\, away from the coordinates
adopted here. \\

BHR\,71-MM & The 2MASS source is 4\arcsec\, away from the coordinates adopted
here. \\

IRAS\,12553-7651 & The 2MASS source is 7\arcsec\, away from the coordinates
adopted here. \\

IRAS\,15398-3359 & The measurements in the 120\arcsec\, apertures are excluded.
The 2MASS source is 3\arcsec\, away from the coordinates adopted here. \\

IRAS\,16017-3936 & The 2MASS source is 9\arcsec\, away from the coordinates
adopted here. \\

VLA\,1623 & This source is a double object consisting of the two sources
VLA\,1623\,A and B, which are comparably bright in the millimeter range. The
measurements from Pudritz et al. \cite{1996ApJ...470L.123P} were excluded. \\

IRAS\,16293-2422 & IRAS\,16293-2422 consits of two sources, IRAS\,16293\,A and
B, seperated by about 4\arcsec\, (Looney et al. \cite{2000ApJ...529..477L}). We
used only the data given for both objects together. \\

IRAS\,16493-4242 & The 2MASS source is 3.5\arcsec\, away from the coordinates
adopted here. \\

IRAS\,18148-0440 & The data taken in the 19 and 120\arcsec\, apertures are
excluded. The 2MASS source is 6\arcsec\, away from the coordinates adopted
here. \\

Serp\,SMM\,5 & The 2MASS source is 4\arcsec\, away from the coordinates adopted
here. \\

Serp\,SMM\,3 & The flux at 3.4\,mm measured in the small aperture was not used.
\\

Serp\,SMM\,2 & The upper limits at long wavelengths are taken within small
apertures and hence do not contradict the fluxes. \\

IRAS\,18331-0035 & The 1.3\,mm flux taken in a 11\arcsec\, aperture was
excluded. The 2MASS source is 3\arcsec\, away from the coordinates adopted
here. \\

HH\,108\,MMS & The 1.3\,mm flux taken in a 11\arcsec\, aperture was excluded.
\\

G\,34.24+0.13\,MM & The 2MASS source is 5\arcsec\, away from the coordinates
adopted here. \\

IRAS\,19156+1906 & The large flux at 1.3\,mm given by Cabrit \& Andr\'e
\cite{1991ApJ...379L..25C} was excluded. Also all data taken in apertures
larger than 100\arcsec, as well as the other large flux taken at 1.3\,mm were
not used. \\

IRAS\,19180+1114 & The 2MASS source is 3\arcsec\, away from the coordinates
adopted here. \\

R\,CrA\,R1 & The 2MASS source is 5\arcsec\, away from the coordinates adopted
here. \\

IRAS\,19345+0727 & IRAS\,19345+0727 shows a large scatter in the sub-mm fluxes.
We did not use the data taken in apertures larger than 120\arcsec. \\

IRAS\,20050+2720 & The 2MASS source is 2.5\arcsec\, away from the coordinates
adopted here. \\

IRAS\,20386+6751 & The sub-mm data taken in apertures larger than 60\arcsec\,
were excluded. There is large scatter in the sub-mm data. \\

IRAS\,21017+6742 & Sub-mm fluxes measured in the 120\arcsec\, apertures are
excluded. There is a large scatter in the sub-mm data. \\

CB\,230 & The 2MASS source is 2\arcsec\, away from the coordinates adopted
here. \\

L\,944-SMM\,1 & The sub-mm fluxes measured in the 50\arcsec\, apertures by
Visser et al. \cite{2002AJ....124.2756V} are used. \\

IC\,1396\,W & The IRAS data might be confused by other cold dust in the
vicinity. \\

CB\,232 & This is a double source consisting of CB\,232\,SMM\,1 and SMM\,2.
SMM\,1 is much brighter in the sub-mm. Hence, all data are attributed to this
object. The 2MASS source is 6\arcsec\, away from the coordinates adopted here.
\\

NGC\,7129\,FIRS\,2 & From the 1.3\,mm points we chose the 1.49\,Jy, since all
the others contradict significantly with the data taken at nearby wavelengths. 
\\

L\,1251\,B & Data taken in the 120\arcsec\, apertures are excluded. \\

L\,1211 & These are several (at least four) sources. Also there are no sub-mm
data. The ISO data represent the object MMS\,4, while the IRAS source might
include all objects (MMS\,1, 2, 3, and 4). The ISO data were used and the IRAS
points are excluded. The 2MASS source is 6\arcsec\, away from the coordinates
adopted here. \\

Cep\,E-MM & This is a double source with a separation of 1.4\arcsec. The flux
of Moro-Mart\'{i}n et al. \cite{2001ApJ...555..146M} at 1.3\,mm is not used due
to the small aperture. \\

NGC\,7538\,S & The 2MASS source is 3\arcsec\, away from the coordinates adopted
here. \\

L\,1246-SMM\,1 & The sub-mm fluxes taken in the 50\arcsec\, apertures by Visser
et al. \cite{2002AJ....124.2756V} are used. \\

IRAS\,23238+7401 & The 120\arcsec\, aperture measurments are excluded. \\

\noalign{\smallskip}
\end{longtable}
\renewcommand{\arraystretch}{1.0}
\end{center}
\vspace{-0.75cm}
$^a$ Object name used in this paper. Further common names are given in
Table\,\ref{sources}. \\
\twocolumn


\end{appendix}

\label{lastpage}


\begin{thebibliography}{}
\bibitem[1999]{1999ApJ...513L..57A}
Andr\'e, P., Motte, F., Bacmann, A., 1999, ApJ, 513L, 57

\bibitem[1993]{1993ApJ...406..122A}
Andr\'e, P., Ward-Thompson, D., Barsony, M., 1993, ApJ, 406, 122

\bibitem[2000]{2000prpl.conf...59A}
Andr\'e, P., Ward-Thompson, D., Barsony, M., 2000, in Protostars and Planets IV, 59 

\bibitem[2002]{2002RMxAA..38...13A}
Anglada, G., Rodr\'{i}guez, L.F., 2002, RMxAA, 38, 13

\bibitem[1992]{1992ApJ...384L..53B}
Barsony, M., Kenyon, S.J., 1992, ApJ, 384L, 53

\bibitem[2001]{2001A&A...372..173B}
Bontemps, S., Andr\'e, P., Kaas, A.A., Nordh, L., Olofsson, G., Huldtgren, M., Abergel, A., et al., 2001, A\&A, 372, 173

\bibitem[1996]{1996A&A...311..858B}
Bontemps, S., Andr\'e, P., Terebey, S., Cabrit, S., 1996, A\&A, 311, 858

\bibitem[1995]{1995A&A...297...98B}
Bontemps, S., Andr\'e, P., Ward-Thompson, D., 1995, A\&A, 297, 98

\bibitem[1997]{1997ApJ...476..781B}
Bourke, T.L., Garay, G., Lehtinen, K.K., Koehnenkamp, I., Launhardt, R., Nyman, L-A., May, J., et al. 1997, ApJ, 476, 781

\bibitem[1991]{1991ApJ...379L..25C}
Cabrit, S., Andr\'e, P., 1991, 379, 25

\bibitem[2000]{2000prpl.conf..377C}
Calvet, N., Hartmann, L., Strom, S.E., 2000, in Protostars and Planets IV, 377

\bibitem[1992]{1992A&A...262..106C}
Carballo, R., Wesselius, P.R., Whittet, D.C.B., 1992, A\&A, 262, 106

\bibitem[1998]{1998Sci...282..462C}
Cernicharo, J., Lefloch, B., Cox, P.; Cesarsky, D., Esteban, C., Yusef-Zadeh, F., Mendez, D. I., et al. 1998, Science, 282, 462

\bibitem[2000]{2000ApJ...530..851C}
Chandler, C.J., Richer, J.S., 2000, ApJ, 530, 851

\bibitem[1997]{1997ApJ...478..295C}
Chen, H., Grenfell, T.G., Myers, P.C., Hughes, J.D., 1997, ApJ, 478, 295

\bibitem[1995a]{1995ApJ...445..377C}
Chen, H., Myers, P.C., Ladd, E.F., Wood, D.O.S., 1995a, ApJ, 445, 377

\bibitem[1997a]{1997ApJ...474L.135C}
Chini, R., Reipurth, B., Ward-Thompson, D., Bally, J., Nyman, L.-A., Sievers, A., Billawala, Y., 1997a, ApJ, 474L, 135

\bibitem[1997b]{1997A&A...325..542C}
Chini, R., Reipurth, B., Sievers, A., Ward-Thompson, D., Haslam, C.G.T., Kreysa, E., Lemke, R., 1997b, A\&A, 325, 542

\bibitem[2001]{2001A&A...369..155C}
Chini, R., Ward-Thompson, D., Kirk, J.M., Nielbock, M., Reipurth, B., Sievers, A., 2001, A\&A, 369, 155

\bibitem[1991]{1991ApJS...75..611C}
Clark, F.O., 1991, ApJS, 75, 611

\bibitem[1997]{1997MNRAS.291..337C}
Codella, C., Muders, D., 1997, MNRAS, 291, 337

\bibitem[1985]{1985ApJ...297..751D}
Dame, T.M., Thaddeus, P., 1985, ApJ, 297, 751

\bibitem[1999]{1999MNRAS.309..141D}
Davis, C.J., Matthews, H.E., Ray, T.P., Dent, W.R.F., Richer, J.S., 1999, MNRAS, 309, 141

\bibitem[1998]{1998MNRAS.301.1049D}
Dent, W.R.F., Matthews, H.E., Ward-Thompson, D., 1998, MNRAS, 301, 1049

\bibitem[1999]{1999AJ....118..972D}
Devine, D., Reipurth, B., Bally, J., 1999, AJ, 118, 972

\bibitem[1998]{1998A&A...335..243E}
Eiroa, C., Palacios, J., Casali, M.M., 1998, A\&A, 335, 243

\bibitem[2003]{2003ApJ...595..259E}
Eisl\"offel, J., Froebrich, D., Stanke, T., McCaughrean, M.J., 2003, 595, 259

\bibitem[2004]{2004A&A.inprep.F}
Froebrich, D., Schmeja, S., Smith, M.D., Klessen, R.S., 2004, A\&A, in prep.

\bibitem[2003]{2003MNRAS.346..163F}
Froebrich, D., Smith, M.D., Hodapp, K.W., Eisl\"offel, J., 2003, MNRAS, in press

\bibitem[2001]{2001A&A...366..873F}
Fuente, A., Neri, R., Mart\'{i}n-Pintado, J., Bachiller, R., Rodr\'{i}guez-Franco, A., Palla, F., 2001, A\&A, 366, 873

\bibitem[2001]{2001ApJ...555...40G}
Giannini, T., Nisini, B., Lorenzetti, D., 2001, ApJ, 555, 40

\bibitem[1998]{1998MNRAS.298..644G}
Gibb, A.G., Davis, C.J., 1998, MNRAS, 298, 644

\bibitem[2000]{2000MNRAS.313..663G}
Gibb, A. G., Little, L.T., 2000, MNRAS, 313, 663

\bibitem[1994]{1994coun.conf..169G}
Güsten, R., 1994, in The Cold Universe, 169

\bibitem[2002]{2002ApJ...578..914H}
Hartmann, L., 2002, ApJ, 578, 914

\bibitem[1997]{1997A&A...323..549H}
Henriksen, R., Andr\'e, P., Bontemps, S., 1997, A\&A, 323, 549

\bibitem[1996]{1996A&A...309..267H}
Holland, W.S., Greaves, J.S., Ward-Thompson, D., Andr\'e, P., 1996, A\&A, 309, 267

\bibitem[1999]{1999ApJ...526..833H}
Huard, T.L., Sandell, G., Weintraub, D.A., 1999, ApJ, 526, 833

\bibitem[1998]{1998ApJ...493L..97H}
Hunter, T.R., Neugebauer, G., Benford, D.J., Matthews, K., Lis, D.C., Serabyn, E., Phillips, T.G., 1998, ApJ, 493L, 97

\bibitem[1996]{1996ApJ...460L..45H}
Hurt, R.L., Barsony, M., 1996, ApJ, 460, 45

\bibitem[2001]{2001ApJ...559..307J}
Johnstone, D., Fich, M., Mitchell, G.F., Moriarty-Schieven, G., 2001, ApJ, 559, 307

\bibitem[2000]{2000A&A...354.1036K}
Kalenskii, S.V., Promislov, V.G., Alakoz, A., Winnberg, A.V., Johansson,  L.E.B., 2000, A\&A, 354, 1036

\bibitem[2001]{2001ApJ...550L..77K}
Klessen, R.S., 2001, ApJ, 550, 77

\bibitem[1993]{1993A&A...272..235K}
Kun, M., Prusti, T., 1993, A\&A, 272, 235

\bibitem[1991a]{1991ApJ...366..203L}
Ladd, E.F., Adams, F.C., Casey, S., Davidson, J.A., Fuller, G.A., Harper, D.A., Myers, P.C., Padman, R., 1991a, ApJ, 366, 203

\bibitem[2003]{l03}
Larson, R.B., 2003, astro-ph, 0306595

\bibitem[1997]{1997A&A...326..329L}
Launhardt, R., Henning, T., 1997, A\&A, 326, 329

\bibitem[1996]{1996A&A...312..569L}
Launhardt, R., Mezger, P.G., Haslam, C.G.T., Kreysa, E., Lemke, R., Sievers, A., Zylka, R., 1996, A\&A, 312, 569

\bibitem[2001]{2001A&A...367..311L}
Lehtinen, K., Haikala, L.K., Mattila, K., Lemke, D., 2001, A\&A, 367, 311

\bibitem[2000]{2000ApJ...529..477L}
Looney, L.W., Mundy, L.G., Welch, W.J., 2000, ApJ, 529, 477

\bibitem[1997]{1997ApJ...489..719M}
Mardones, D., Myers, P.C., Tafalla, M., Wilner, D.J., Bachiller, R., Garay, G., 1997, ApJ, 489, 719

\bibitem[1992a]{1992A&A...256..631M}
Mezger, P.G., Sievers, A.W., Haslam, C.G.T., Kreysa, E., Lemke, R., Mauersberger, R., Wilson, T.L., 1992a, A\&A, 256, 631

\bibitem[1998]{1998ApJ...505L..39M}
Molinari, S., Testi, L., Brand, J., Cesaroni, R., Pallo, F., 1998, ApJ, 505L, 39

\bibitem[1994]{1994ApJ...436..800M}
Moriarty-Schieven, G.H., Wannier, P.G., Keene, J., Tamura, M., 1994, ApJ, 436, 800

\bibitem[2001]{2001ApJ...555..146M}
Moro-Mart\'{i}n, A., Noriega-Crespo, A., Molinari, S., Testi, L., Cernicharo, J., Sargent, A., 2001, ApJ, 555, 146

\bibitem[2001]{2001A&A...365..440M}
Motte, F., Andr\'e, P., 2001, A\&A, 365, 440

\bibitem[1998]{1998ApJ...492..703M}
Myers, P.C., Adams, F.C., Chen, H., Schaff, E., 1998, ApJ, 492, 703

\bibitem[2002]{2002A&A...393.1035N}
Nisini, B., Caratti o Garatti, A., Giannini, T., Lorenzetti, D., 2002, A\&A, 393, 1035

\bibitem[1999]{1999ApJ...515..696O}
O'Linger, J., Wolf-Chase, G., Barsony, M., Ward-Thompson, D., 1999, ApJ, 515, 696

\bibitem[1997]{1997A&A...326..373O}
Olmi, L., Felli, M., Cesaroni, R., 1997, A\&A, 326, 373

\bibitem[1999]{1999ApJ...520..223P}
Park, Y.-S., Kim, J., Minh, Y.C., 1999, ApJ, 520, 223

\bibitem[1994]{1994A&A...282..233P}
Persi, P., Ferrari-Toniolo, M., Marenzi, A.R., Anglada, G., Chini, R., Kr\"ugel, E., Sep\'ulveda, I., 1994, A\&A, 282, 233

\bibitem[2001]{2001A&A...376..907P}
Persi, P., Marenzi, A.R., G\'omez, M., Olofsson, G., 2001, A\&A, 376, 907

\bibitem[2001]{2001MNRAS.326..927P}
Phillips, R.R., Gibb, A.G., Little, L.T., 2001, MNRAS, 326, 927

\bibitem[1992a]{1992A&A...260..151P}
Prusti, T., Whittet, D.C.B., Assendorp, R., Wesselius, P.R., 1992a, A\&A, 260, 151

\bibitem[1996]{1996ApJ...470L.123P}
Pudritz, R.E., Wilson, C.D., Carlstrom, J.E., Lay, O.P., Hills, R.E., Ward-Thompson, D., 1996, ApJ, 470L, 123

\bibitem[2003]{2003AJ....125.2568R}
Rebull, L.M., Cole, D.M., Stapelfeldt, K.R., Werner, M.W., 2003, AJ, 125, 2568

\bibitem[1996]{1996A&A...314..258R}
Reipurth, B., Nyman, L.-A., Chini, R., 1996, A\&A, 314, 258

\bibitem[1999]{1999AJ....118..983R}
Reipurth, B., Rodr\'{i}guez, L.F., Chini, R., 1999, AJ, 118, 983

\bibitem[2004]{2004A&A..inprep.R}
Rengel, M., Eisl\"offel, J., Hodapp, K.W., 2004, A\&A, in prep.

\bibitem[1993]{1993MNRAS.262..839R}
Richer, J.S., Padman, R., Ward-Thompson, D., Hills, R.E., Harris, A.I., 1993, MNRAS, 262, 839

\bibitem[1996]{1996RMxAA..32...27R}
Rodr\'{i}guez, L.F., Reipurth, B., 1996, RMxAA, 32, 27

\bibitem[2003]{2003ApJ...598.1100R}
Rodr\'{i}guez, L.F., G\'omez, Y., Reipurth, B., ApJ, 598, 1100

\bibitem[1999]{1999ApJ...519..236S}
Sandell, G., Avery, L.W., Baas, F., Coulson, I., Dent, W.R.F., Friberg, P., Gear, W.P.K., Greaves, J., Holland, W., Jenness, T., et al., 1999, ApJ, 519, 236

\bibitem[2001]{2001ApJ...546L..49S}
Sandell, G., Knee, L.B.G., 2001, ApJ, 546L, 49

\bibitem[2003]{2003ApJ...590L..45S}
Sandell, G., Wright, M., Forster, J.R., 2003, ApJ, 590, 45

\bibitem[1996]{1996A&A...309..827S}
Saraceno, P., Andr\'e, P., Ceccarelli, C., Griffin, M., Molinari, S., 1996, A\&A, 309, 827

\bibitem[2004]{2004A&A...419..405S}
Schmeja, S., Klessen, R.S., 2004, A\&A, 419, 405

\bibitem[2000]{2000ApJS..131..249S}
Shirley, Y.L., Evans II, N.J., Rawlings, J.M.C., Gregersen, E.M., 2000, ApJS, 131, 249

\bibitem[2004]{2004ApJ...601..930S}
Shu, F.H., Li, Z.-Y., Allen, A., 2004, ApJ, 601, 930

\bibitem[1998]{1998Ap&SS.261..169S} 
Smith M.D., 1998, Ap\&SS, 261, 169

\bibitem[2000]{2000IrAJ...27...25S}
Smith, M.D., 2000, IrAJ, 27, 25

\bibitem[2002]{s02}
Smith, M.D., 2002, The Origins of Stars and Planets: The VLT View, J.Alves \& M.McCaughrean (ed.)

\bibitem[1999]{1999A&A...348..479T}
Tafalla, M., Myers, P.C., Mardones, D., Bachiller, R., 1999, A\&A, 348, 479

\bibitem[2003]{2003A&A...407..237T}
Thompson, M.A., Macdonald, G.H. , 2003, A\&A, 407, 237

\bibitem[2002]{2002AJ....124.2756V}
Visser, A.E., Richer, J.S., Chandler, C.J., 2002, AJ, 124, 2756

\bibitem[2001]{2001A&A...379..208V}
Vuong, M.H., Cambr\'esy, L., Epchtein, N., 2001, A\&A, 379, 208

\bibitem[1981]{1981A&A....97..329W}
Wamsteker, W., 1981, A\&A, 97, 329

\bibitem[2001]{2001MNRAS.327..955W}
Ward-Thompson, D., Buckley, H.D., 2001, MNRAS, 327, 955

\bibitem[1995b]{1995MNRAS.273L..25W}
Ward-Thompson, D., Eiroa, C., Casali, M.M., 1995b, MNRAS, 273L, 25

\bibitem[1995]{1995ApJ...449L..73W}
Wilner, D.J., Welch, W.J., Forster, J.R., 1995, ApJ, 449L, 73

\bibitem[2003]{2003MNRAS.344..809W}
Wolf-Chase, G., Moriarty-Schieven, G., Fich, M., Barsony, M., 2003, MNRAS, 344, 809

\bibitem[2003]{2003ApJS..145..111Y}
Young, C.H., Shirley, Y.L., Evans II, N.J., Rawlings, J.M.C., 2003, ApJS, 145, 111

\bibitem[1992]{1992A&A...265..726Z}
Zinnecker, H., Bastien, P., Arcoragi, J.-P., Yorke, H.W., 1992, A\&A, 265, 726

\end{thebibliography}
\end{document}